\documentclass[10pt,journal,compsoc]{IEEEtran}
\usepackage{graphicx}
\usepackage{url}
\usepackage{subcaption}
\usepackage{amsmath}
\usepackage{algorithm}
\usepackage{algcompatible}

\usepackage{multirow}
\usepackage[table,xcdraw]{xcolor}
\usepackage{color}

\usepackage{caption} 
\usepackage{tabularx}

\ifCLASSOPTIONcompsoc
\else
\fi

\ifCLASSINFOpdf
\else
\fi

\begin{document}

\title{Cloud Service Provider Evaluation System using Fuzzy Rough Set Technique}

\author{Parwat Singh Anjana$^{1}$, Priyanka Badiwal$^{2}$, Rajeev Wankar$^{2}$, and C. Raghavendra Rao$^{2}$
{cs17resch11004@iith.ac.in, \{priyanka.badiwal, rajeev.wankar, crrcs.ailab\}@gmail.com}

\IEEEcompsocitemizethanks{\IEEEcompsocthanksitem Parwat Singh Anjana is with the $^{1}$Department of CSE, Indian Institute of Technology Hyderabad, Hyderabad, India - 502285.\protect\\
% note need leading \protect in front of \\ to get a newline within \thanks as
% \\ is fragile and will error, could use \hfil\break instead.
E-mail: cs17resch11004@iith.ac.in
\IEEEcompsocthanksitem Priyanka Badiwal, Rajeev Wankar, and C. Raghavendra Rao are with $^{2}$School of Computer and Information Sciences, Hyderabad Central University, Hyderabad, India - 500019. Email: \{priyanka.badiwal, rajeev.wankar, crrcs.ailab\}@gmail.com}% <-this % stops a space
\thanks{}}

%% The paper headers
\markboth{Preprint submitted to IEEE Transactions on Cloud Computing}%
{Parwat \MakeLowercase{\textit{et al.}}: CSP Evaluation System for Ranking and Selection using FRST}

\IEEEcompsoctitleabstractindextext{%
\begin{abstract}
Cloud Service Providers (CSPs) offer a wide variety of scalable, flexible, and cost-efficient services to cloud users on demand and pay-per-utilization basis. However, vast diversity in available services leads to various challenges to determine and select the best suitable service for a user. Also, sometimes user need to hire the required services from multiple CSPs which introduce interface, account, security, support, and Service Level Agreements (SLAs) management difficulties. To circumvent such problems having a Cloud Service Broker (CSB) be aware of service offerings and user Quality of Service (QoS) requirements will benefit both the CSPs as well as a user. In this work, we proposed a Fuzzy Rough Set based Cloud Service Brokerage Architecture, which is responsible for ranking and selecting services based on users QoS requirements, and finally monitor the service execution. We have used fuzzy rough set technique for dimension reduction. Used weighted Euclidean distance to rank the CSPs. To prioritize user QoS request, we intended to use user assign weights, also incorporated system assigned weights to give the relative importance to QoS attributes. We compared the proposed ranking technique with an existing method based on the system response time. The case study experiment result shows that the proposed approach is scalable, resilience, and produce better results with less searching time.

\end{abstract}

%\keywords{Cloud Service Provider (CSP), Cloud Service Broker (CSB), Fuzzy Rough Set Theory (FRST), Reduct, Quality of Service (QoS), Service Provider Ranking, Service Provider Selection.}

}

% make the title area
\maketitle

\IEEEdisplaynotcompsoctitleabstractindextext

\IEEEpeerreviewmaketitle

\section{Introduction}
\label{Introduction}
\IEEEPARstart{W}{ith} the emergence of Cloud, Cloud Service Providers (CSPs) offers a wide variety of flexible, scalable, on-demand, and pay-as go online resources to the cloud users \cite{garg2013framework}. Nowadays, Cloud becomes an indispensable component of the many organizations; it requires a right approach for adoption, deployment, and preservation of the resources \cite{rane2015cloud}. Also, it introduces several challenges for the cloud user for choosing the best CSP among a considerable number of CSPs \cite{Anjana2017}. Further, organizations may even need to rent different services from different CSPs which lead to the challenges of operating multiple interfaces, accounts, supports, and Service Level Agreements (SLAs) \cite{rane2015cloud}. To facilitate the CSPs and to circumvent the difficulties faced by an organization (while for ranking, selecting, and dealing with multiple CSPs) we need a Cloud Service Broker (CSB). According to Gartner \cite{csb}, a CSB is a third party (individual or an organization) between CSPs and users. It provides intermediation, aggregation, and arbitrage services to consult, mediate, and facilitates the cloud computing solutions on behalf of the users or business organizations \cite{csb}. An \textbf{\textit{Intermediation}} service broker offers additional value-added service on top of existing services by appending specific capabilities, for example, identity, security, and access reporting and administration. To advance data transfer security and integration, an \textit{\textbf{Aggregation}} service broker includes consolidation and integration of various services, for example, data integration and portability assurance between different CSPs and users. An \textit{\textbf{Arbitrage}} service broker is similar to aggregation broker, but the number of services being aggregated will not be fixed. An Arbitrage service broker acquires comprehensive services and presents it in smaller containers with more excellent cumulative value to the users, for example, a large block of bandwidth at wholesale rates. A CSB reduces processing costs, increase flexibility, provides access to various CSPs within one contract channel, and offers timely execution of services by removing acquisition limits through reporting and billing policies \cite{guzek2015cloud}. The proposed architecture is an \textit{\textbf{Aggregation}} service broker.

Recent advancement in CSB focuses on developing efficient systems and strategies that help the user to select and monitor cloud resource efficiently. Service execution and evaluation help in getting historical information about services and affected by dynamic, quantifiable and non-quantifiable QoS attribute \cite{buyya2009cloud}. A quantifiable QoS attribute mainly refers to functional QoS requirements and can efficiently be measured without any ambiguity, e.g., vCPU speed (frequency), service response time, cost, etc. While a non-quantifiable attribute primarily depends on non-functional QoS requirements and cannot be quantified easily. In general, a non-quantifiable attribute depends on user experience, e.g., accountability, security, feedback, support, etc. \cite{garg2013framework}. However, the importance of quantifiable and non-quantifiable attributes are classified, existing approaches do not present an efficient technique to handle non-quantifiable QoS attributes efficiently and objectively. Furthermore, quantified patterns are used to analyze user QoS requirements in existing techniques \cite{garg2013framework} \cite{qu2014cloud}. While user requirements are imprecise or vague, we need a technique to handle it efficiently. The precise QoS attribute comprises only crisp values (un-ambiguous) while imprecise QoS attribute usually includes fuzzy values that cannot be quantified. We presented a hybrid imprecision technique which consists of quantifiable and non-quantifiable QoS attributes. A service user submits his desired QoS requirement along with weights using a Graphical User Interface (GUI) during CSP ranking phase.

In the proposed architecture information system consists both quantifiable and non-quantifiable QoS attributes (to describe CSP service offerings and to formulate user's QoS requirements). We employ Fuzzy Rough Set Technique (FRST) to deal with hybrid information system with real-valued entries (to provide the solution for real-time conditional attributes with a fuzzy decision) and for search space reduction. For search space reduction all reducts of the decision system are computed using FRST, and the best reduct is selected to generate Reduced Decision System (RDS). The best reduct is the reduct which consists maximum overlapping QoS attributes with user QoS request. The proposed architecture is an \textit{Aggregation} broker designed using FRST that offers the cloud user facility to rank, select, and monitor the services based on their desired QoS requirements.

The major contribution of this work includes: 
\begin{itemize}
    \item An effective cloud service broker to rank service providers based on the user QoS requirements, to select and monitor the service execution of the selected service provider.
    \item The principal focus is on dealing with hybrid system and search space reduction using fuzzy rough set technique. And to improve the accuracy of service provider selection with the incorporation of dynamic and network layer QoS parameters (dynamic and Network QoS parameter changes with time and availability of the resources).
    \item To oblige the CSPs to give satisfying services to compete in further assignments by induction of user experience (feedback) along with CSPs performance monitoring during execution. By using this factor (past performances, user experience) in the service description, we enhance the correctness of ranking procedure.
    \item Introduction to user assigned weights to support user to prioritize their needs along with system assigned weights to QoS attributes during the ranking procedure to improve ranking efficiency. Subsequently allowing the user to choose their desired CSP from a ranked list.
\end{itemize}

\par The rest of the paper is structured in five sections as follows. The detailed study of existing work and significant contributions to CSP selection (using rough, fuzzy set) presented in $Section$ \ref{relatedwork}. $Section$ \ref{frst} provides the overview of Fuzzy Rough Set and need of Fuzzy Rough Set Approach. $Section$ \ref{proposedarch} introduce proposed architecture and basic blocks of the architecture along with ranking algorithm. $Section$ \ref{casestudy} gives a comprehensive case study on ranking compute cloud service providers (Infrastructure as a Service) along with results. At the end $Section$ \ref{conclusion}, conclude with some future directions.

\section{Related Work}
\label{relatedwork}
With the advancement and increasing use of cloud services, researchers analyzed the CSPs for various types of application. A wide range of discovery and selection techniques have been developed for the evaluation of CSPs based on QoS requirements of the user. This section presents the work carried out by researchers for CSP ranking and selection using rough, fuzzy set theory based techniques, along with some other significant contribution which includes essential specification used in this paper.

To address the challenges of CSPs discovery and selection, Le et al., \cite{sun2014cloud} presented a comprehensive survey of existing service selection approaches. They evaluate the CSP selection approaches based on five aspects and characterized into four groups (Multi-Criteria Decision Making, Multi-Criteria Optimization, Logic-Based, and Other Approaches). The Multi-Criteria Decision-Making based approaches have successfully implemented to discover desired cloud services. It includes Analytic Hierarchy Process (AHP), Multi-Attribute Utility Theory (MAUT), Outranking, and Simple Additive Weighting based techniques \cite{sun2014cloud} which are an extension of Web services. Godse et al., \cite{godse2009approach} presented an MCDM based cloud services selection technique, performed a case study to prove the significance of methodology to solve SaaS selection. Garg et al., \cite{garg2013framework} introduced AHP technique based CSP ranking framework known as SMICloud. This framework enables users to compare and select the CSP using ``Service Measurement Index (SMI)'' attributes to satisfy users QoS requirements. They computed key performance indicators defined in the ``Cloud Service Measurement Index Consortium \cite{CSMIC} QoS standards to compare cloud services. However, they did not examine trustworthiness, user experience, and Network Layer QoS attribute for ranking and selection of CSPs.

Alhamad et al., \cite{alhamad2011trust} introduced a fuzzy set theory based technique for CSP selection using availability, security, usability, and scalability as QoS attributes. Hussain et al., \cite{ur2012iaas} proposed an MCDM based CSP selection technique for IaaS services and presented a case study on service provider selection among thirteen providers using five performance criteria. A Cloud-QoS Management Strategy (C-QoSMS) using rough set technique was proposed by Ganghishetti et al., \cite{ganghishetti2011quality}. Specifically, they considered ranking of IaaS cloud services using SLA and QoS attributes. They also extended their work in Modified C-QoSMS \cite{ganghishetti2011rough} and presented a case study using  Random and Round Robin algorithms. However, they did not examine non-quantifiable QoS attribute and considered only categorical information hence they need to discretize the numerical value for selection of cloud services. Qu et al., \cite{qu2014cloud} introduced a fuzzy hierarchy based trust evaluation using inference system to evaluates users trust based on fuzzy QoS requirements of users and progressive fulfillment of services to advance CSP selection. With a case study and simulation, they illustrated the effectiveness and efficiency of their model. In \cite{patiniotakis2015pulsar} Patiniotakis et al., introduced a PuLSaR: Preference-based Cloud Service Recommender system which is an MCDM based optimization brokerage technique, in the proposed method to deal vagueness of imprecise QoS attributes they used fuzzy set theory. Furthermore, to demonstrate the performance of the proposed procedure, they conducted experiments. In \cite{aruna2016framework} Aruna et al., suggested a fuzzy set based ranking framework for federated cloud IaaS infrastructure to rank the CSPs based on QoS attributes and SLAs. Their proposed framework consists of three phases of AHP process as decomposition of the problem, priority judgment, and aggregation with simple rule inductions. The first contribution to Fuzzy Rough Set based CSP ranking is introduced by Anjana et al., \cite{Anjana2017}. They proposed a Fuzzy Rough Set Based Cloud Service Broker (FRSCB) architecture in which they did the QoS Attribute Categorization into different types, also includes network layer and non-quantifiable QoS attributes, ranked the CSPs by mean of the total score using Euclidean distance. They presented a case study with fifteen service provider along with fifteen SMI based QoS attributes.

The proposed work E-FRSCB is an extension of work presented in FRSCB architecture \cite{Anjana2017}. In our work, to deal with dynamic quantifiable, and non-quantifiable characteristics of service measurement index based QoS attributes, we used the fuzzy rough set based hybrid technique. We assigned different weights to attributes at different levels of ranking procedure, incorporate quantifiable and non-quantifiable QoS attributes including network layer parameters, fetch real-time values of dynamic attributes, monitor service execution to improve next ranking assignment. We also simulated the behavior of our proposed work using CloudSim \cite{cloudsim} and demonstrated the significance of E-FRSCB against FRSCB.

\section{Fuzzy Rough Set Theory (FRST)}
\label{frst}
In traditional set theory, human reasoning is described using a Boolean logic i.e. true or false (0/1) and are not enough to reason efficiently. Therefore there is a need for decision terms that take the value ranging within an interval from 0 to 1 to represent human reasoning. The Fuzzy Set Theory (FST) proposed by Lotfi Zadeh \cite{zadeh1965fuzzy} in 1965 can realize human reasoning in the form of a degree 'd' such that 0 $<=$ d $<=$ 1. For example, in FST a person is healthy by 70\% (d = $0.7$) or unhealthy by 30\% (d = $0.3$), while in traditional set theory a person can be healthy (1/true) or unhealthy (0/false). FST based set membership function determined by the Equation \ref{eq:eq1}.
\begin{equation}\label{eq:eq1}
    \mu_y(X)\in [0,1]
\end{equation}
 Where: $y \in X$ i.e. $y$ is an element of $X$, and $X$ is a set of elements.

Rough Set Theory (RST) proposed by Pawlak \cite{pawlak2007rudiments} is a mathematical way to deal with vagueness (uncertainty) present in the data. RST proves its importance in Artificial Intelligence manly in expert systems, decision systems, knowledge discovery, acquisition, and many more. The fundamental advantage of using RST is that there is no need for holding any prior knowledge or information about data. The notion of boundary region is used to describe the uncertainty associated with data in RST. A set is defined as a rough set when boundary region is non-empty while defined as a crisp set when boundary region is empty \cite{pawlak2007rudiments}. 

One limitation of the RST is that it can deal with only categorical or non-quantifiable attributes. In general, quantifiable and interval-fuzzy values also exist in real-world data as explained in Section \ref{Introduction}. The RST fails when we have quantifiable data in our Information System (IS) (TABLE \ref{infsys})). One plausible solution to the problem can be obtained by performing discretion so that quantifiable attributes can be categorized and FST can be employed. Alternatively, we can also use FST to deal directly with quantifiable characteristics in the IS. However to deal with a hybrid information system (as shown in the TABLE \ref{infsys}) which consists of both categorical (non-quantifiable) and quantifiable attribute we need a hybrid technique. Therefore, the Fuzzy Rough Set Theory (FRST) can be employed for CSP ranking and selection. A FRST is a generalized form of a crisp RST and FST, it is derived from the approximation of a fuzzy set in a crisp approximation space. FRST is helpful when we are dealing with a decision system in with conditional attributes are real-valued attributes \cite{chen2012novel}. It is primarily employed in the search space reduction to improve the classification accuracy in several aspects including storage, accuracy, speed \cite{chen2012novel}. Search space reduction can be achieved by determining reduct of the system. A reduct is a minimal subset of attributes of the system that give the same classification power as given by the entire set of attributes \cite{jensen2008rough}. In a real-valued (conditional attributes) based decision system, it is done with the help of finding a minimum set of conditional attribute that preserves discernment information by concerning decision attribute. For further detailed understanding of discernibility matrix based all reduct computation of FRST (which we used in this paper), readers can refer to \cite{package_rst}, \cite{chen2012novel}. 

\par The proposed FRST based hybrid technique deals with hybrid real-valued information system and also provides the solution for real-time conditional attributes with the fuzzy decision. It computes all possible reducts of the decision system (TABLE \ref{DS}) using FRST all reduct computation function presented in \cite{package_rst}, and selected the best reduct using \textit{Best Reduct Algorithm (Algorithm \ref{algo:BestReduct})} for search space reduction. It incorporates the user feedback and monitors the cloud service execution using Service Execution Monitor (Section \ref{SEM}) once CSP selection is made by the users to improve the accuracy of further CSP selection. Which is missing in most of the existing CSB techniques.

\section{Proposed Brokerage Architecture}
\label{proposedarch}
\subsection{System Architecture}
\begin{figure}[htb]
    \hspace{-.5cm}
     {\includegraphics[width=.37\textheight, height=.28\textheight]{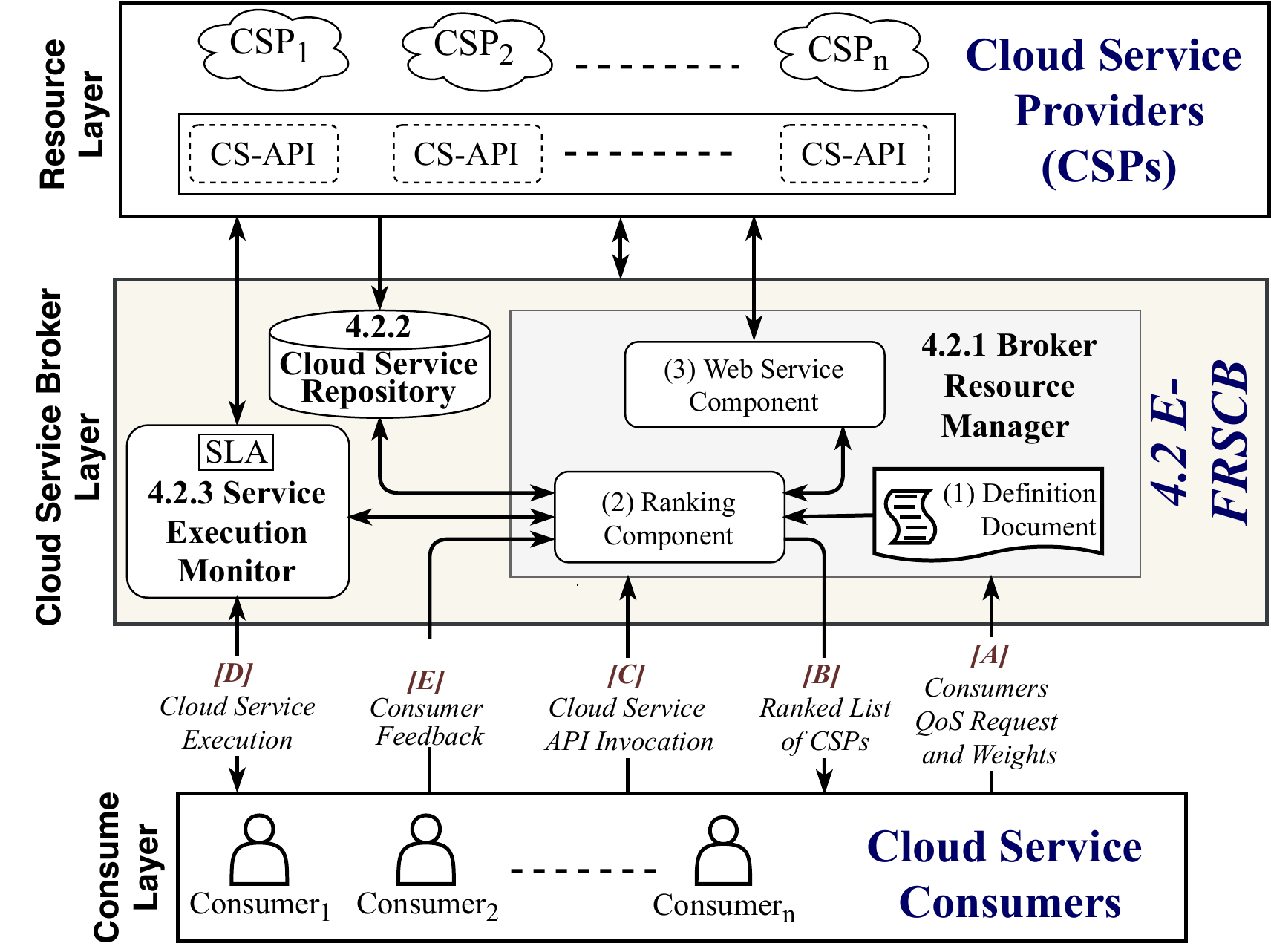}}
    \caption{Extended Fuzzy Rough Set based Cloud Service Brokerage Architecture. (Square Bracket shows E-FRSCB algorithm steps)}
   \label{fig:EFRSCB}
\end{figure}

The proposed architecture attempts to help the cloud users by providing cloud brokerage services such as ranking CSPs, selection of the best CSP, after selection execution, and monitoring of service execution. The proposed Extended-Fuzzy Rough Set based Cloud Service Broker (E-FRSCB) brokerage architecture (Figure \ref{fig:EFRSCB}) consists of several basic software components classified into three layers as Cloud User Layer, Cloud Service Broker Layer, and Resource Layer. Cloud User User layer includes the number of cloud users either requesting for service provider ranking or using cloud services. Cloud Service Broker (CSB) layer is the central component of the architecture responsible for CSPs ranking, selection, service executions, and monitoring (we focused on this layer only). It consists of Cloud Service Repository (CSR), Service Execution Monitor (SEM) and Broker Resource Manager (BRM). Finally, Resource layer includes the number of CSPs along with service models which is practiced using a simulator (CloudSim) \cite{cloudsim}. A cloud user requesting for brokering services at time `t' with QoS request and attribute weight is stored in an individual service definition document with BRM. The detailed introduction to each component of E-FRSCB is introduced in Subsection \ref{components}.

\begin{figure}[htb]
    \hspace{-.2cm}
     {\includegraphics[width=.37\textheight, height=.28\textheight]{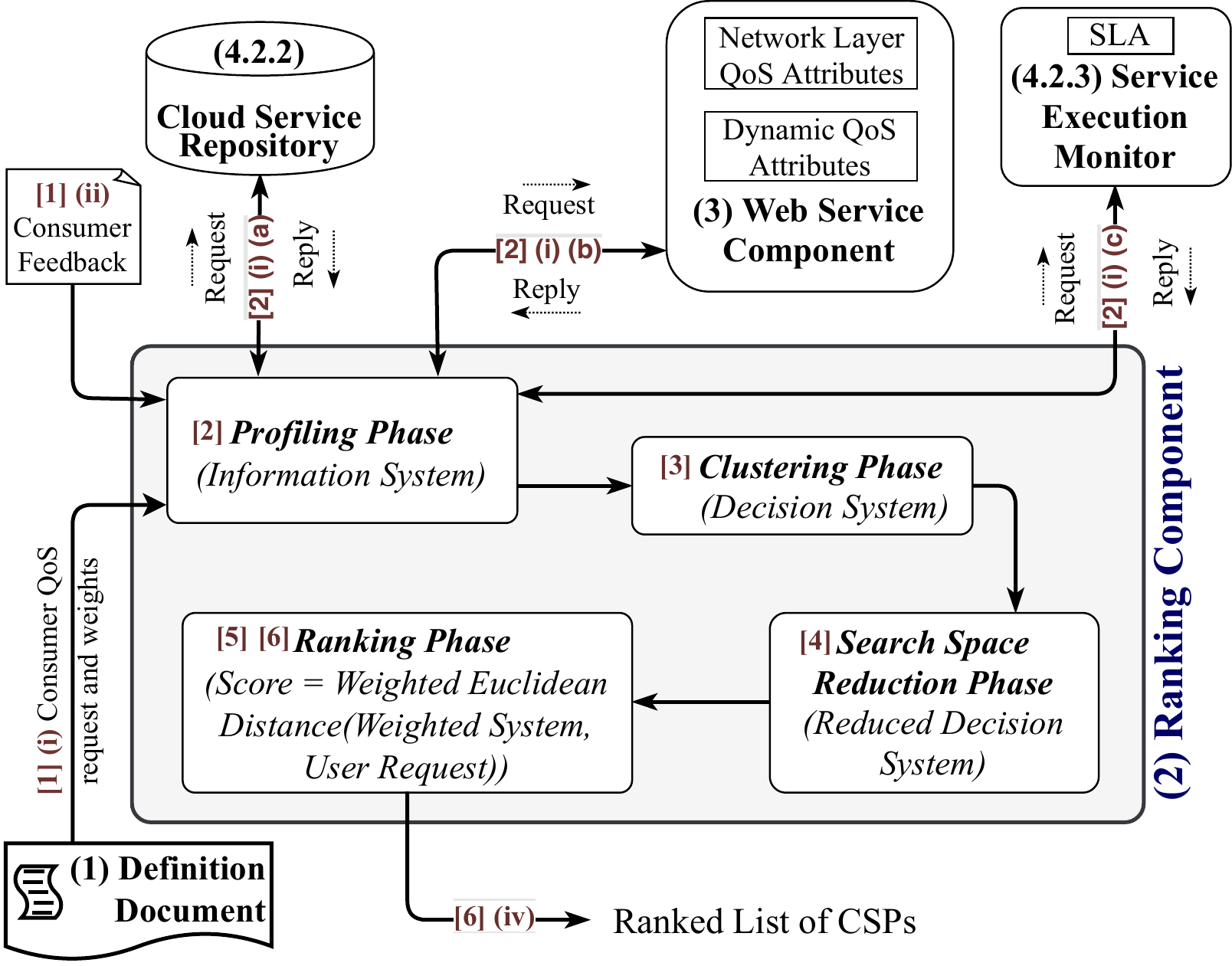}}
    \caption{Service Ranking Procedure a High-level View (Information-flow)}
   \label{fig:RC}
\end{figure}

\begin{algorithm*}[ht]
\small 
\caption{E-FRSCB Algorithm}
{\textbf{Input}}{: Definition Document, Feedback, CSPs Service Information}\\
{\textbf{Output}}{: Ranked List of Cloud Service Providers (CSPs), Service Execution}
\label{algo:E-FRSCB}
 \begin{algorithmic}[]
 \STATE \textbf{procedure} {e-frscb}{$(user_i$)}\\
 
  \STATE STEP [A] (i) {E-FRSCB }{$\gets$}{ User-Request(QoS values, weights)}\\
  
  \STATE STEP [B] (i) {Definition Document }{$\gets$}{ STEP [A] (i)}\\ \hspace{1.6cm}
                  (ii) {Ranked CSPs List }{$\gets$}{ Ranking$\_$Algorithm(DD, CF, SEM, WSC, CSR)}\\
          
  \STATE STEP [C] (i) {User-ID }{$\gets$}{ STEP [B] (ii)} \\ \hspace{1.59cm}
                (ii) User-Select one CSP, send CSP-ID to E-FRSCB \\ \hspace{1.59cm}
                (iii) {User-invoke(CS-API) }{$\gets$}{ Selected(CSP)}\\ \hspace{1.59cm}
                (iv) {E-FRSCB-SLA-User-ID }{$\gets$}{ establish-SLA(User-ID, CSP-ID)}\\

  \STATE STEP [D] (i) E-FRSCB-BRM-Resource$\_$Reservation(User-ID, Service API)\\ \hspace{1.63cm}
                 (ii) E-FRSCB-BRM-Service$\_$Exe(User-ID, Service API, SLA-User-ID)\\
  \STATE STEP [E] (i) {Profile-CSP-ID }{$\gets$}{ User-ID-Feedback}\\
\end{algorithmic}

%\tiny
{
$/*$ BRM: Broker Resource Manager; CSR: Cloud Service Registry; DD: Definition Document; CF: User Feedback; SEM: Service Execution Monitor; SLA: Service Level Agreement; DS: Decision System; DQoS: Dynamic QoS Attributes; NLQoS: Network Layer QoS Attributes; CSP: Cloud Service Provider.
$*/$}
\end{algorithm*}

\subsection{E-FRSCB Components} \label{components}
\subsubsection{Broker Resource Manager (BRM):}
\label{BRM}
The overall functionality of our proposed system is controlled by Broker Resource Manager (BRM) from ranking CSPs to service execution monitoring. Figure \ref{fig:EFRSCB} shows BRM as a principal component of E-FRSCB architecture. It has several sub-components such as \textit{Cloud Service Repository (CSR), Ranking Component (RC), Web Service Component (WSC),} and \textit{Service Execution Monitor (SEM)}. The profile of service providers are accumulated in the information container know as \textit{Cloud Service Repository}. \textit{Ranking Component} takes user QoS request, performs CSP ranking, and returns the ranked list as shown in Figure \ref{fig:RC}. On every new user request, \textit{Web Service Component} is used to fetch the real-time values of dynamic QoS attributes including network layer QoS parameters. A user submits his QoS request using GUI, on submission \textit{E-FRSCB Algorithm} (Algorithm \ref{algo:E-FRSCB}) is invoked. Once service selection is made by user, E-FRSCB algorithm contacts the selected service provider. The SLA will be established between user and CSP via negotiation process. Furthermore, resources will be reserved, and service execution will be performed. Finally, \textit{Service Execution Monitor} monitors the execution of the services and maintain an \textit{Execution Monitoring Log} to keep track of the CSPs performance and to facilitate next CSP ranking process.

\textbf{(1) Definition Document:} Definition document is used to record the user QoS request along with desire priority given by the user in the form of weights to each requested QoS attribute (TABLE \ref{tbl:cqos}). The Definition document is utilized in ranking and service execution monitoring phase.

\textbf{(2) Ranking Component:} Ranking begins with invocation of Ranking Algorithm (Algorithm \ref{algo:ranking}) of Ranking Component. It is a central part of the BRM and is composed of several phases such as information gathering, analysis, search space reduction, and ranking. For these, it interacts with Cloud Service Repository (CSR), Service Execution Monitor (SEM), and Web Service Components (WSC). 
\begin{itemize}
    \item \textbf{Profiling Phase} does the information gathering by sending a request to CSR, SEM, and WSC. Once profiling phase receives the response from all the part, it generates the Information System (IS) based on the latest information of monitoring phase and dynamic QoS attribute including network layer QoS attributes (using WSC gets the current value of QoS attributes). At the end of this phase, it sends the IS along with user QoS request to next phase (clustering phase) for further information analysis. 
    
    \item At \textbf{Clustering Phase} to designing Decision System (DS) \textit{K-means} \cite{jain2010data} is employed over the IS which gives different clustering labels. Each object of the IS (CSP) is associated with respective clustering label to generate the DS, i.e., each cluster is labeled with distinct labels and labels are used as a decision attribute in DS. During this process, if CSPs are grouped under the same clustering label then they offer related service. In the proposed method to determine the optimal number of clustering labels \textit{Elbow} method \cite{tibshirani2001estimating} is employed, and decision attribute is kept at the end of the DS (i.e., last row) as shown in TABLE \ref{DS}. 
    
    \item \textbf{Search Space Reduction Phase} In this phase, reduct concept of Fuzzy Rough Set Theory (FRST) is applied. All reducts of the DS (TABLE \ref{tbl:reduct}) are computed using \textit{all reduct computation function} presented in \cite{package_rst}, and the best reduct is selected using \textit{Best Reduct Algorithm (Algorithm \ref{algo:BestReduct})}. The best reduct is the reduct which consists the maximum overlapping QoS attributes with user QoS requests. If more then one reducts have the same number of user requested QoS attribute then to break the tie \textit{Best Reduct Algorithm} selects the reduct which has more number of dynamic QoS attributes. Based on the selected reduct Reduced Decision System (RDS) is generated. Finally based on the RDS and Definition Document ranking of CSPs is performed. 
    
    \item \textbf{Ranking Phase}, \textit{Weighted Euclidean Distance (Score)} is computed for CSPs by using attribute values of CSP and user request. Smaller the \textit{Score} represents better CSPs, therefore ranking is done based on increasing order of \textit{Score}. Finally, ranking algorithm terminates with sending ranked list of CSPs to BRM.
    
\end{itemize}

\textbf{\textit{Relative Weight Assignment for QoS Attributes:}} We need to consider the relative importance of QoS attributes for service provider comparison before calculating respective \textit{Score} of the CSPs. For this, we need to assign some weights to QoS attributes. We employed \textit{System Assigned Weights} and \textit{User Assigned Weights} to respective QoS attributes (user requested QoS and Network Layer QoS in RDS TABLE \ref{RDS}). During weight assignment, at the first-level system will assign weights, while at the second-level customer can assign his QoS preferred weights. By doing this, we try to incorporate both user preferences and actual relative importance of quality attributes in the ranking process. Assigning the weight at lower levels prioritize the user request alongside also gives importance to the qualities which are not part of the user request but have a critical role in the ranking process (e.g., network QoS attributes) at the higher level.

\textit{System Assigned Weights:} If user requested attribute is not present in RDS, then that quality does not have enough potential concerning selected reduct and hence not counted for ranking process. However, all other attributes which are not part of the user request and are part of the RDS are assigned the weights based on Golden Section Search Method \cite{zhao1994determination}. Where \textit{66.67}\% (i.e., 0.67) weights allocated to user requested attributes and $33.33$\% (i.e., 0.33) weights to others which are part of the RDS (e.g., Network Layer QoS attributes, monitored qualities). Here the sum of the weight is considered to be equal to 1. A critical remark to acknowledge here is that if the user does not wish to assign weights. Then only one level of weight assignment will be done for the ranking process.
    
\textit{User Assigned Weights:} The user assigned weights indicate the relative importance of QoS attribute sought by the users in their request. A user can use his/her own scale to assign different weights to QoS attributes in his request \cite{garg2013framework} (as shown in TABLE \ref{tblweights}). This weight assignment is done based on the suggestion given in AHP technique \cite{saaty2005theory}, for this, the restriction of the sum of all weights need not be equal to one, unlike system assigned weights. However, for each first level user requested QoS attribute (e.g., Accountability, Agility, Cost, Assurance, Security, and Satisfaction) we considered the sum of the weights is equal to 1 as shown in TABLE \ref{tbl:cqos}. This weight assignment technique is proposed originally to assign different weights to each QoS attributes in AHP technique which we adopted for weight assignment in our proposed E-FRSCB architecture.

\begin{table}[ht]
\centering
\caption{Relative Importance of QoS Attributes}
%\vspace{-.13cm}
\label{tblweights}
\resizebox{.3\textwidth}{!}{%
\begin{tabular}{|c|c|}
\hline
\textbf{Relative Importance} & \textbf{Value} \\ \hline
Equaly important & 1 \\ \hline
Somewhat more important & 3 \\ \hline
Definitely more important & 5 \\ \hline
Much more important & 7 \\ \hline
Extremely more important & 9 \\ \hline
\end{tabular}%
}
\end{table}

\textbf{(3) Web Service Component (WSC):} It is used to fetches the current state of dynamic QoS attributes including network layer QoS attributes from CSPs using web-services. The state of dynamic QoS attributes change from one ranking process to another ranking process, and we need the real-time values to improve the accuracy of the ranking process. For this, we used various APIs (Cloud Harmony APIs \cite{cloudharmony}) to fetch the current state of dynamic QoS attribute of selected CSPs. Dynamic QoS attributes are specified in advance, so there is no need of determining them again during each ranking process.

\subsubsection{Cloud Service Repository (CSR):} It is a repository service employed to store information about CSPs services. CSPs needs to register their QoS plans/services, capabilities, offerings, and initial SLAs in our Cloud Service Repository (CSR). In the proposed architecture, we assume CSPs have been recorded their services in CSR beforehand. So from CSR, we can obtain required information of the CSPs, their QoS service offerings, and other information to generate the initial Information System (IS) as shown in TABLE \ref{infsys}. In the absence of global CSR, the CSPs need to register their services in our local CSR registry. In the Figure \ref{fig:EFRSCB} and Figure \ref{fig:RC} we have shown our local CSR, as for experiment purpose we used local CSR.

\subsubsection{Service Execution Monitor (SEM):}
\label{SEM}
The process of cloud monitoring includes dynamic tracking of the QoS attributes relevant to virtualized cloud resources, for example, vCPU, Storage, Network, etc. \cite{da2016monitoring}. The cloud computing resources configuration is a genuinely challenging task which consists of various heterogeneous virtualized cloud computing resources \cite{alhamazani2015overview}. Furthermore, sometime, there will be a massive demand for a specific cloud service. And because of the change in requirement and availability of various resources including network parameter the performance may change. Which directly and indirectly also affects the user service experience. Therefore, there is a need for cloud monitoring to keep track of resources operation at high demand, to detect fluctuations in performance and to account the SLA breaches of specific QoS attributes \cite{syed2017cloud}. The performance fluctuations may also happen due to failures and other runtime configuration. Cloud monitoring solutions (tools) can be classified into three types as Generic solutions, Cluster-Grid solutions, and Cloud specific solutions. Cloud specific solutions are designed explicitly to monitor computing environments in the cloud and developed by academic researchers or commercial efforts  \cite{da2016monitoring}. Existing Cloud specific solutions (tools) includes Amazon CloudWatch, Private Cloud Monitoring Systems (PCMONS) (open source \cite{PCMON}), Cloud Management System (CMS), Runtime Model for Cloud Monitoring (RMCM), CAdvisor (open source), and Flexible Automated Cloud Monitoring Slices (Flex-ACMS) \cite{da2016monitoring}.

In CSB, cloud service monitoring can be used to improve the CSP ranking and to build healthy competition between CSPs to adhere to provide excellent services to the user to contest in next CSP ranking process. In proposed E-FRSCB execution monitor is implemented using \cite{PCMON}. Furthermore, it can also be implemented as a third party monitoring service offered by CSPs (e.g., Amazon CloudWatch, AppDynamics). Service execution monitor is used to provide the guarantee that the deployed cloud services perform (performance monitoring aspect) at the expected level to satisfy the SLA established between user and CSPs. This component includes monitoring task regarding the currently running services (offered by CSPs selected by respective users). This responsibility consists of the detection and collection of QoS attribute values. Data collected during monitoring is utilized for next CSP ranking process and is send to BRM whenever BRM issues a new request for data.

\begin{algorithm*}[ht]
\small
\caption{Ranking Algorithm}
{\textbf{Input}}{: Definition Document, User Feedback, QoS Attributes, Optimal Number of Clusters}\\
{\textbf{Output}}{: Ranked List of Cloud Service Providers (CSPs)}
\label{algo:ranking}
\begin{algorithmic}[]
%\Procedure{Ranking}{$DD, CF, SEM, WSC, CSR$}\\
\STATE \textbf{procedure} {ranking}{$(definition\_Document, user\_Feedback, qos, clustering\_Labels, optimal\_Clusters)$}
 \STATE STEP [1] (i) {RC }{$\gets$}{ definition\_Document}\\ \hspace{1.59cm}
                (ii) {RC }{$\gets$}{ user\_Feedback}\\
  
  \STATE STEP [2] (i) Fetch latest information about QoS from different components\\ \hspace{1.99cm}
                    (a) {qoS\_Attribute}{ $\gets$ }{request(CSR)}\\ \hspace{1.99cm}
                    (b) {dynamic\_QoS\_Values}{ $\gets$ }{request(WSC)}\\ \hspace{1.99cm}
                    (c) {performance\_QoS}{ $\gets$ }{request(SEM)}\\ \hspace{1.59cm}
                (ii) {IS}{ $\gets$ }{generate([1](ii), [2](i)(a), [2](i)(b), [2](i)(c))}\\

  \STATE STEP [3] (i) {clustering\_Labels}{ $\gets$ }{kmeans(IS, optimal\_Clusters, nstart)}\\ \hspace{1.59cm}
                 (ii) {decision\_ Attribute}{ $\gets$ }{clustering\_Label(IS)}\\ \hspace{1.59cm}
                 (iii) {DS}{ $\gets$ }{generateDS(IS, decision\_Attribute)}\\
  
  \STATE STEP [4] (i) {all\_Reducts}{ $\gets$ }{FS.all.reducts.computation(DS)}\\ \hspace{1.59cm}
  				  (ii) {best\_Reduct}{ $\gets$ }{best.Reduct(DS, definition\_Document)}\\ \hspace{1.59cm}
  				  (iii) {RDS}{ $\gets$ }{generate(DS, best\_Reduct)}\\
                
  \STATE STEP [5] (i) {RDSN}{ $\gets$ }{generate(RDS, NQoS)}\\ \hspace{1.59cm}
				  (ii) Weight Assignment\\ \hspace{1.99cm}
				  		(a) {W\_RDSN}{ $\gets$ }{assign(RDS QoS 67\%, NQoS 33\%)}\\ \hspace{1.99cm}
				  		(b) \textbf{if}(definition\_Document(user.Weights))\\ \hspace{2.92cm}
				  			{W\_RDSN}{ $\gets$ }{assign(W\_RDSN, user.Weights)}\\
  
  \STATE STEP [6] (i) {score-CSP}{ $\gets$ }{weighted\_Euclidean\_Distance(user\_Request, W\_RDSN)} \\ \hspace{1.59cm} 
  				  (ii) if CSPs Score is equal, give priority to CSP with more Dynamic QoS\\ \hspace{1.59cm}
  				  (iii) {ranked\_List\_CSPs}{ $\gets$ }{ascending\_order(score-CSP)}\\ \hspace{1.59cm}
  				  (iv) return(ranked\_List\_CSPs)\\
\end{algorithmic}

%\tiny
{
$/*$ 
RC: Ranking Component; IS: Information System; DS: Decision System; RDS: Reduced Decision System; NQoS:  Network QoS Attributes; RDSN: Reduced Decision System + Network QoS Attributes; W\_RDSN: Weighted RDSN;
$*/$}
\end{algorithm*}

\begin{algorithm*}[ht]
\small 
\caption{Best Reduct Algorithm}
{\textbf{Input}}{: Decision System, All Reducts of Decision System, Definition Document}\\
{\textbf{Output}}{: Best Reduct}
\label{algo:BestReduct}
 \begin{algorithmic}[]
% \Procedure{BestReduct}{$DD, CF, SEM, WSC, CSR$}\Comment{}\\
 \STATE \textbf{procedure} {best.Reduct}{$(Decision\_System, All\_Reducts, Definition\_Document)$}
  \STATE STEP [A] \textit{find} number of overlapping QoS with User QoS Request for Each Reduct\\
  \STATE STEP [B] \textit{select} all the Reduct which has maximum number of overlapping QoS\\
  \STATE STEP [C] \textit{if} more then one such Reduct selected which have maximum overlapping QoS\\ \hspace{1.7cm}
                (i) \textit{count} Number of Dynamic QoS in each such Reduct\\ \hspace{1.7cm}
                (ii) \textit{select} the Reduct which has more number of Dynamic QoS\\\hspace{1.7cm}
                (iii) \textit{if} more then one such Reduct has equal number of Dynamic QoS \textit{select} anyone Reduct\\
  \STATE STEP [D] \textit{return}(selected.Reduct)          
\end{algorithmic}
\end{algorithm*}

\section{CASE STUDY: Compute Cloud Service Provider (IaaS CSP) Ranking based on User QoS Requirement}
\label{casestudy}
The ranking method of E-FRSCB Architecture given in the Section \ref{proposedarch} is analyzed for Computation Services (IaaS)  offered by CSPs with the help of a case study example in this section. However, this can also work with other types of services like SaaS, PaaS also. $RStudio$ \cite{rstudio} is used as a development IDE and $R-language$ \cite{team2000r} for implementation. To submit the user QoS request to the system GUI is developed using $fgui$ \cite{fgui} package. We have referred Cloud Service Measurement Index Consortium (CSMIC) \cite{CSMIC}, defined Service Measurement Index (SMI) matrices for evaluation of Compute Cloud Service Providers and to design initial Information System (IS) (TABLE \ref{infsys}) for this we considered $10$ CSPs along with total $17$ QoS Attributes (scalable). We have designed IS for general purpose Compute Cloud Service Provider (IaaS) services by considering $8$ first level SMI matrices which consist of $17$ third level QoS attributes. We experimented with synthesized data, however, tried to incorporate actual QoS values. We have taken data from different sources including CSPs websites, SLAs, literature \cite{garg2013framework}, and CloudSim \cite{cloudsim} for most of the QoS as shown in TABLE \ref{infsys}. For dynamic and Network Layer QoS attributes (Availability, Latency, Throughput, Service Downtime) data is collected using Cloud Harmony API \cite{cloudharmony}. For performance-intensive QoS attribute (vCPU speed, Response Time) information from Service Execution Monitor is used. However, in initial IS design, actual values of vCPU speed offered by CSPs are used, and for Response Time values are assigned randomly. The categorical (non-quantifiable) QoS attribute (such as security, accountability, user feedback, and supports) values are randomly assigned. System assigned weights to QoS attribute are assigned using Golden Section Search Method \cite{zhao1994determination}, while the user specifies desired weights along with QoS request (as explained in the Section \ref{BRM}).

In the literature to standardize the categorical (qualitative/non-quantifiable) QoS attributes for CSPs ranking, there is no single globally accepted standardization. The categorical attribute represents the enumerated type of quality where the expected value of the attribute is presented in the form of levels \cite{gould2015introductory}. In proposed architecture for categorical attributes such as accountability, support, security, and user feedback different levels ([1:10]) are introduced based on the work presented in \cite{Anjana2017}. We demonstrate here the quantification of Security Levels in proposed technique (for support type, accountability, and user feedback level quantification is done similarly).

To rank the CSPs security is one of the critical Cloud QoS metrics; its primary objective is to strengthen the security mechanisms and mitigate threats. It helps in developing the user trust and improves the operational performance. In E-FRSCB, security consists the ten different level where random values are assigned for all the CSPs. It includes various performance indicators such as certificates provisioning, vCPU configuration, firewall, access and log management policies, encryption/masking of data, etc. Assurance framework proposed by European Union Network Information Security Agency (ENISA) for Cloud consists of 10, 69, 130 first, second, and third-degree security indicators respectively \cite{catteddu2009cloud}. Cloud Security Alliance introduced fundamental principles that service providers must follow and support user for security estimation \cite{CSA}. We did security levels quantification into ten different levels [1:10] (can be easily extended) based on the risk of security threat and security performance indicators. In quantization, level 1 represents the most straightforward security mechanisms while level 10 represent the complex and highest level of security mechanisms offered by CSPs. Each level of quantized security level consists of one or more security performance indices (130 third-degree security indicators). The straightforward example can be at level 1 only provider and user authentication is done, at level 2 including level 1 multi-factor and fine granular authentication and authorization is performed. Similarly, at further levels (i.e., from level 3-10) including upper-level different firewall administration, privileged controls over accesses, application identification and other security metrics takes place to achieve higher security. In the following, we present the proposed ranking method in multiple steps.

In the first step, whenever a new user submits his QoS request (TABLE \ref{tbl:cqos}), Ranking Algorithm  (Algorithm \ref{algo:ranking}) is invoked. During this process to generate the initial IS, it fetches the data from CSR for Compute Cloud Services (IaaS) offered by CSPs, as shown in TABLE \ref{infsys}. To fetch the actual values of dynamic and network layer QoS attribute ranking algorithm send a request to WSC component to execute the web services.

\begin{table*}[ht]
\centering
\caption{Information System (IS). row: attributes of IS; column: objects of IS}
\label{infsys}
%\hspace{-1cm}
\resizebox{\textwidth}{!}{%
\begin{tabular}{|c|c|c|c|c|c|c|c|c|c|c|c|c|c|c|}
\hline
\multicolumn{2}{|c|}{\multirow{2}{*}{\textbf{QoS Attributes}}} & \multirow{2}{*}{\textbf{\begin{tabular}[c]{@{}c@{}}QoS\\ Unit\end{tabular}}} & \multicolumn{10}{c|}{\textbf{Cloud Service Providers (CSPs)}} & \multirow{2}{*}{\textbf{\begin{tabular}[c]{@{}c@{}}QoS\\ Type\end{tabular}}} & \multirow{2}{*}{\textbf{\begin{tabular}[c]{@{}c@{}}Require-\\ ment\end{tabular}}} \\ \cline{4-13}
\multicolumn{2}{|c|}{} &  & \textbf{\begin{tabular}[c]{@{}c@{}}Google\\ Compute\\ Engine\end{tabular}} & \textbf{\begin{tabular}[c]{@{}c@{}}Storm\\ on\\ Demand\end{tabular}} & \textbf{\begin{tabular}[c]{@{}c@{}}Century\\ Link\end{tabular}} & \textbf{\begin{tabular}[c]{@{}c@{}}Amazon\\ EC2\end{tabular}} & \textbf{\begin{tabular}[c]{@{}c@{}}Vultr\\ Cloud\end{tabular}} & \textbf{\begin{tabular}[c]{@{}c@{}}IBM\\ Soft\\ Layer\end{tabular}} & \textbf{Linode} & \textbf{\begin{tabular}[c]{@{}c@{}}Digital\\ Ocean\end{tabular}} & \textbf{\begin{tabular}[c]{@{}c@{}}Micro-\\ soft\\ Azure\end{tabular}} & \textbf{\begin{tabular}[c]{@{}c@{}}Racks-\\ pace\end{tabular}} &  &  \\ \hline
\textbf{Accountability} & \textbf{Levels} & \textbf{(1-10)} & 8 & 4 & 4 & 9 & 3 & 7 & 1 & 6 & 8 & 2 & \multirow{7}{*}{\textbf{Dynamic}} & \textbf{Categorical} \\ \cline{1-13} \cline{15-15} 
\multirow{6}{*}{\textbf{\begin{tabular}[c]{@{}c@{}}Agility\\ (Capacity)\end{tabular}}} & \multirow{3}{*}{\textbf{\begin{tabular}[c]{@{}c@{}}Number\\ of\\ vCPUs\end{tabular}}} & \multirow{3}{*}{\textbf{\begin{tabular}[c]{@{}c@{}}(4 core\\ each)\end{tabular}}} & \multirow{3}{*}{16} & \multirow{3}{*}{8} & \multirow{3}{*}{1} & \multirow{3}{*}{8} & \multirow{3}{*}{6} & \multirow{3}{*}{4} & \multirow{3}{*}{3} & \multirow{3}{*}{8} & \multirow{3}{*}{8} & \multirow{3}{*}{8} &  & \multirow{12}{*}{\textbf{Numerical}} \\
 &  &  &  &  &  &  &  &  &  &  &  &  &  &  \\
 &  &  &  &  &  &  &  &  &  &  &  &  &  &  \\ \cline{2-13}
 & \textbf{vCPU Speed} & \textbf{(GHZ)} & 2.6 & 2.7 & 3.6 & 3.6 & 3.8 & 3.4 & 2.3 & 2.2 & 3.5 & 3.3 &  &  \\ \cline{2-13}
 & \textbf{Disk} & \textbf{(TB)} & 3 & 0.89 & 2 & 1 & 2 & 1 & 0.76 & 2 & 0.195 & 1 &  &  \\ \cline{2-13}
 & \textbf{Memory} & \textbf{(GB)} & 14.4 & 16 & 16 & 15 & 24 & 32 & 16 & 16 & 32 & 15 &  &  \\ \cline{1-14}
\multirow{6}{*}{\textbf{Cost}} & \textbf{vCPU} & \textbf{(\$/h)} & 0.56 & 0.48 & 0.56 & 0.39 & 0.44 & 0.69 & 0.48 & 0.23 & 0.38 & 0.51 & \multirow{8}{*}{\textbf{Static}} &  \\ \cline{2-13}
 & \multirow{4}{*}{\textbf{\begin{tabular}[c]{@{}c@{}}Data\\ Transfer\\ Bandwidth\end{tabular}}} & \multirow{2}{*}{\textbf{\begin{tabular}[c]{@{}c@{}}In\\ (\$/TB-m)\end{tabular}}} & \multirow{2}{*}{0} & \multirow{2}{*}{8} & \multirow{2}{*}{10} & \multirow{2}{*}{0} & \multirow{2}{*}{7} & \multirow{2}{*}{8} & \multirow{2}{*}{10} & \multirow{2}{*}{9} & \multirow{2}{*}{18.43} & \multirow{2}{*}{15.2} &  &  \\
 &  &  &  &  &  &  &  &  &  &  &  &  &  &  \\ \cline{3-13}
 &  & \multirow{2}{*}{\textbf{\begin{tabular}[c]{@{}c@{}}Out\\ (\$/TB-m)\end{tabular}}} & \multirow{2}{*}{70.6} & \multirow{2}{*}{40} & \multirow{2}{*}{51.2} & \multirow{2}{*}{51.2} & \multirow{2}{*}{22.86} & \multirow{2}{*}{92.16} & \multirow{2}{*}{51.2} & \multirow{2}{*}{45.72} & \multirow{2}{*}{18.43} & \multirow{2}{*}{40} &  &  \\
 &  &  &  &  &  &  &  &  &  &  &  &  &  &  \\ \cline{2-13}
 & \textbf{Storage} & \textbf{(\$/TB-m)} & 40.96 & 122.8 & 40.96 & 21.5 & 102 & 36.86 & 80.86 & 40.96 & 40.96 & 36.86 &  &  \\ \cline{1-13} \cline{15-15} 
\multirow{3}{*}{\textbf{Assurance}} & \multirow{2}{*}{\textbf{\begin{tabular}[c]{@{}c@{}}Support\\ (Levels)\end{tabular}}} & \multirow{2}{*}{\textbf{(1-10)}} & \multirow{2}{*}{8} & \multirow{2}{*}{4} & \multirow{2}{*}{7} & \multirow{2}{*}{10} & \multirow{2}{*}{5} & \multirow{2}{*}{7} & \multirow{2}{*}{3} & \multirow{2}{*}{10} & \multirow{2}{*}{8} & \multirow{2}{*}{2} &  & \multirow{2}{*}{\textbf{Categorical}} \\
 &  &  &  &  &  &  &  &  &  &  &  &  &  &  \\ \cline{2-15} 
 & \textbf{Availability} & \textbf{(\%)} & 99.99 & 100 & 99.97 & 99.99 & 99.89 & 99.97 & 100 & 99.99 & 100 & 99.95 & \textbf{Dynamic} & \textbf{Numerical} \\ \hline
\textbf{Security} & \textbf{Levels} & \textbf{(1-10)} & 9 & 8 & 6 & 10 & 8 & 10 & 6 & 8 & 10 & 2 & \textbf{Static} & \multirow{3}{*}{\textbf{Categorical}} \\ \cline{1-14}
\multirow{2}{*}{\textbf{Satisfaction}} & \multirow{2}{*}{\textbf{\begin{tabular}[c]{@{}c@{}}Feedback\\ (Levels)\end{tabular}}} & \multirow{2}{*}{\textbf{(1-10)}} & \multirow{2}{*}{9} & \multirow{2}{*}{7} & \multirow{2}{*}{6} & \multirow{2}{*}{10} & \multirow{2}{*}{6} & \multirow{2}{*}{9} & \multirow{2}{*}{6} & \multirow{2}{*}{8} & \multirow{2}{*}{9} & \multirow{2}{*}{7} & \multirow{10}{*}{\textbf{Dynamic}} &  \\
 &  &  &  &  &  &  &  &  &  &  &  &  &  &  \\ \cline{1-13} \cline{15-15} 
\multirow{4}{*}{\textbf{Performance}} & \multirow{2}{*}{\textbf{\begin{tabular}[c]{@{}c@{}}Response\\ Time\end{tabular}}} & \multirow{2}{*}{\textbf{(sec)}} & \multirow{2}{*}{83.5} & \multirow{2}{*}{90} & \multirow{2}{*}{97} & \multirow{2}{*}{52} & \multirow{2}{*}{90} & \multirow{2}{*}{100} & \multirow{2}{*}{97} & \multirow{2}{*}{85} & \multirow{2}{*}{76} & \multirow{2}{*}{57} &  & \multirow{8}{*}{\textbf{Numerical}} \\
 &  &  &  &  &  &  &  &  &  &  &  &  &  &  \\ \cline{2-13}
 & \multirow{2}{*}{\textbf{vCPU Speed}} & \multirow{2}{*}{\textbf{(GHZ)}} & \multirow{2}{*}{2.6} & \multirow{2}{*}{2.7} & \multirow{2}{*}{3.6} & \multirow{2}{*}{3.6} & \multirow{2}{*}{3.8} & \multirow{2}{*}{3.4} & \multirow{2}{*}{2.3} & \multirow{2}{*}{2.2} & \multirow{2}{*}{3.5} & \multirow{2}{*}{3.3} &  &  \\
 &  &  &  &  &  &  &  &  &  &  &  &  &  &  \\ \cline{1-13}
\multirow{4}{*}{\textbf{\begin{tabular}[c]{@{}c@{}}Network\\ Layer\\ QoS\end{tabular}}} & \multirow{2}{*}{\textbf{\begin{tabular}[c]{@{}c@{}}Down\\ Time\end{tabular}}} & \multirow{2}{*}{\textbf{(min)}} & \multirow{2}{*}{1.02} & \multirow{2}{*}{0} & \multirow{2}{*}{1.98} & \multirow{2}{*}{0.51} & \multirow{2}{*}{9.05} & \multirow{2}{*}{7.5} & \multirow{2}{*}{0} & \multirow{2}{*}{1.53} & \multirow{2}{*}{2.83} & \multirow{2}{*}{2.83} &  &  \\
 &  &  &  &  &  &  &  &  &  &  &  &  &  &  \\ \cline{2-13}
 & \textbf{Latency} & \textbf{(ms)} & 31 & 57 & 31 & 29 & 28 & 29 & 28 & 28 & 32 & 32 &  &  \\ \cline{2-13}
 & \textbf{Throughput} & \textbf{(mb/s)} & 20.24 & 16.99 & 24.99 & 16.23 & 10.11 & 16.23 & 8.12 & 24.67 & 23.11 & 23.67 &  &  \\ \hline
\end{tabular}%
}
\end{table*}

In the second step, on generated IS, \textit{K-means Clustering} (optimal number of clusters (i.e., k value) is determined using Elbow Method \cite{tibshirani2001estimating}) is performed. Where generated clustering labels are used as a decision attribute to design the Decision System (DS) and corresponding clustering label is attached to the CSPs. In the paper, for presentation clarity, we transferred the tables (IS, DS,  Reduced Decision System, and Ranking Table) where row shows attribute and column shows the object of the IS/DS. Therefore clustering labels are kept in the last row as shown in TABLE \ref{DS}.

\begin{table*}[ht]
\centering
\caption{Decision System (DS)}
\label{DS}
\resizebox{\textwidth}{!}{%
\begin{tabular}{|c|c|c|c|c|c|c|c|c|c|c|c|c|}
\hline
\multicolumn{2}{|c|}{} &  & \multicolumn{10}{c|}{\textbf{Cloud Service Providers (CSPs)}} \\ \cline{4-13} 
\multicolumn{2}{|c|}{\multirow{-2}{*}{\textbf{QoS Attributes}}} & \multirow{-2}{*}{\textbf{\begin{tabular}[c]{@{}c@{}}QoS\\ Unit\end{tabular}}} & \textbf{\begin{tabular}[c]{@{}c@{}}Google\\ Compute\\ Engine\end{tabular}} & \textbf{\begin{tabular}[c]{@{}c@{}}Storm\\ on\\ Demand\end{tabular}} & \textbf{\begin{tabular}[c]{@{}c@{}}Century\\ Link\end{tabular}} & \textbf{\begin{tabular}[c]{@{}c@{}}Amazon\\ EC2\end{tabular}} & \textbf{\begin{tabular}[c]{@{}c@{}}Vultr\\ Cloud\end{tabular}} & \textbf{\begin{tabular}[c]{@{}c@{}}IBM\\ Soft\\ Layer\end{tabular}} & \textbf{Linode} & \textbf{\begin{tabular}[c]{@{}c@{}}Digital\\ Ocean\end{tabular}} & \textbf{\begin{tabular}[c]{@{}c@{}}Micro-\\ soft\\ Azure\end{tabular}} & \textbf{\begin{tabular}[c]{@{}c@{}}Racks-\\ pace\end{tabular}} \\ \hline
\textbf{Accountability} & \textbf{Levels} & \textbf{(1-10)} & 8 & 4 & 4 & 9 & 3 & 7 & 1 & 6 & 8 & 2 \\ \hline
 &  &  &  &  &  &  &  &  &  &  &  &  \\
 &  &  &  &  &  &  &  &  &  &  &  &  \\
 & \multirow{-3}{*}{\textbf{\begin{tabular}[c]{@{}c@{}}Number\\ of\\ vCPUs\end{tabular}}} & \multirow{-3}{*}{\textbf{\begin{tabular}[c]{@{}c@{}}(4-core\\ each)\end{tabular}}} & \multirow{-3}{*}{16} & \multirow{-3}{*}{8} & \multirow{-3}{*}{1} & \multirow{-3}{*}{8} & \multirow{-3}{*}{6} & \multirow{-3}{*}{4} & \multirow{-3}{*}{3} & \multirow{-3}{*}{8} & \multirow{-3}{*}{8} & \multirow{-3}{*}{8} \\ \cline{2-13} 
 & \textbf{vCPU Speed} & \textbf{(GHZ)} & 2.6 & 2.7 & 3.6 & 3.6 & 3.8 & 3.4 & 2.3 & 2.2 & 3.5 & 3.3 \\ \cline{2-13} 
 & \textbf{Disk} & \textbf{(TB)} & 3 & 0.89 & 2 & 1 & 2 & 1 & 0.76 & 2 & 0.195 & 1 \\ \cline{2-13} 
\multirow{-6}{*}{\textbf{\begin{tabular}[c]{@{}c@{}}Agility\\ (Capacity)\end{tabular}}} & \textbf{Memory} & \textbf{(GB)} & 14.4 & 16 & 16 & 15 & 24 & 32 & 16 & 16 & 32 & 15 \\ \hline
 & \textbf{vCPU} & \textbf{(\$/h)} & 0.56 & 0.48 & 0.56 & 0.39 & 0.44 & 0.69 & 0.48 & 0.23 & 0.38 & 0.51 \\ \cline{2-13} 
 &  &  &  &  &  &  &  &  &  &  &  &  \\
 &  & \multirow{-2}{*}{\textbf{\begin{tabular}[c]{@{}c@{}}In\\ (\$/TB-m)\end{tabular}}} & \multirow{-2}{*}{0} & \multirow{-2}{*}{8} & \multirow{-2}{*}{10} & \multirow{-2}{*}{0} & \multirow{-2}{*}{7} & \multirow{-2}{*}{8} & \multirow{-2}{*}{10} & \multirow{-2}{*}{9} & \multirow{-2}{*}{18.43} & \multirow{-2}{*}{15.2} \\ \cline{3-13} 
 &  &  &  &  &  &  &  &  &  &  &  &  \\
 & \multirow{-4}{*}{\textbf{\begin{tabular}[c]{@{}c@{}}Data\\ Transfer\\ Bandwidth\end{tabular}}} & \multirow{-2}{*}{\textbf{\begin{tabular}[c]{@{}c@{}}Out\\ (\$/TB-m)\end{tabular}}} & \multirow{-2}{*}{70.6} & \multirow{-2}{*}{40} & \multirow{-2}{*}{51.2} & \multirow{-2}{*}{51.2} & \multirow{-2}{*}{22.86} & \multirow{-2}{*}{92.16} & \multirow{-2}{*}{51.2} & \multirow{-2}{*}{45.72} & \multirow{-2}{*}{18.43} & \multirow{-2}{*}{40} \\ \cline{2-13} 
\multirow{-6}{*}{\textbf{Cost}} & \textbf{Storage} & \textbf{(\$/TB-m)} & 40.96 & 122.8 & 40.96 & 21.5 & 102 & 36.86 & 80.86 & 40.96 & 40.96 & 36.86 \\ \hline
 &  &  &  &  &  &  &  &  &  &  &  &  \\
 & \multirow{-2}{*}{\textbf{\begin{tabular}[c]{@{}c@{}}Support\\ (Levels)\end{tabular}}} & \multirow{-2}{*}{\textbf{(1-10)}} & \multirow{-2}{*}{8} & \multirow{-2}{*}{4} & \multirow{-2}{*}{7} & \multirow{-2}{*}{10} & \multirow{-2}{*}{5} & \multirow{-2}{*}{7} & \multirow{-2}{*}{3} & \multirow{-2}{*}{10} & \multirow{-2}{*}{8} & \multirow{-2}{*}{2} \\ \cline{2-13} 
\multirow{-3}{*}{\textbf{Assurance}} & \textbf{Availability} & \textbf{(\%)} & 99.99 & 100 & 99.97 & 99.99 & 99.89 & 99.97 & 100 & 99.99 & 100 & 99.95 \\ \hline
\textbf{Security} & \textbf{Levels} & \textbf{(1-10)} & 9 & 8 & 6 & 10 & 8 & 10 & 6 & 8 & 10 & 2 \\ \hline
 &  &  &  &  &  &  &  &  &  &  &  &  \\
\multirow{-2}{*}{\textbf{Satisfaction}} & \multirow{-2}{*}{\textbf{\begin{tabular}[c]{@{}c@{}}Feedback\\ (Levels)\end{tabular}}} & \multirow{-2}{*}{\textbf{(1-10)}} & \multirow{-2}{*}{9} & \multirow{-2}{*}{7} & \multirow{-2}{*}{6} & \multirow{-2}{*}{10} & \multirow{-2}{*}{6} & \multirow{-2}{*}{9} & \multirow{-2}{*}{6} & \multirow{-2}{*}{8} & \multirow{-2}{*}{9} & \multirow{-2}{*}{7} \\ \hline
 &  &  &  &  &  &  &  &  &  &  &  &  \\
 & \multirow{-2}{*}{\textbf{\begin{tabular}[c]{@{}c@{}}Response\\ Time\end{tabular}}} & \multirow{-2}{*}{\textbf{(sec)}} & \multirow{-2}{*}{83.5} & \multirow{-2}{*}{90} & \multirow{-2}{*}{97} & \multirow{-2}{*}{52} & \multirow{-2}{*}{90} & \multirow{-2}{*}{100} & \multirow{-2}{*}{97} & \multirow{-2}{*}{85} & \multirow{-2}{*}{76} & \multirow{-2}{*}{57} \\ \cline{2-13} 
 &  &  &  &  &  &  &  &  &  &  &  &  \\
\multirow{-4}{*}{\textbf{Performance}} & \multirow{-2}{*}{\textbf{vCPU Speed}} & \multirow{-2}{*}{\textbf{(GHZ)}} & \multirow{-2}{*}{2.6} & \multirow{-2}{*}{2.7} & \multirow{-2}{*}{3.6} & \multirow{-2}{*}{3.6} & \multirow{-2}{*}{3.8} & \multirow{-2}{*}{3.4} & \multirow{-2}{*}{2.3} & \multirow{-2}{*}{2.2} & \multirow{-2}{*}{3.5} & \multirow{-2}{*}{3.3} \\ \hline
 &  &  &  &  &  &  &  &  &  &  &  &  \\
 & \multirow{-2}{*}{\textbf{\begin{tabular}[c]{@{}c@{}}Down\\ Time\end{tabular}}} & \multirow{-2}{*}{\textbf{(min)}} & \multirow{-2}{*}{1.02} & \multirow{-2}{*}{0} & \multirow{-2}{*}{1.98} & \multirow{-2}{*}{0.51} & \multirow{-2}{*}{9.05} & \multirow{-2}{*}{7.5} & \multirow{-2}{*}{0} & \multirow{-2}{*}{1.53} & \multirow{-2}{*}{2.83} & \multirow{-2}{*}{2.83} \\ \cline{2-13} 
 & \textbf{Latency} & \textbf{(ms)} & 31 & 57 & 31 & 29 & 28 & 29 & 28 & 28 & 32 & 32 \\ \cline{2-13} 
\multirow{-4}{*}{\textbf{\begin{tabular}[c]{@{}c@{}}Network\\ Layer\\ QoS\end{tabular}}} & \textbf{Throughput} & \textbf{(mb/s)} & 20.24 & 16.99 & 24.99 & 16.23 & 10.11 & 16.23 & 8.12 & 24.67 & 23.11 & 23.67 \\ \hline
\multicolumn{3}{|c|}{{\color[HTML]{9A0000} \textbf{Decision Attribute}}} & {\color[HTML]{680100} \textbf{3}} & {\color[HTML]{680100} \textbf{1}} & {\color[HTML]{680100} \textbf{3}} & {\color[HTML]{680100} \textbf{2}} & {\color[HTML]{680100} \textbf{1}} & {\color[HTML]{680100} \textbf{3}} & {\color[HTML]{680100} \textbf{1}} & {\color[HTML]{680100} \textbf{2}} & {\color[HTML]{680100} \textbf{2}} & {\color[HTML]{680100} \textbf{2}} \\ \hline
\end{tabular}%
}
\end{table*}

\begin{table}[ht]
\centering
\caption{Definition Document: User QoS Request and Weights}
\label{tbl:cqos}
\resizebox{.45\textwidth}{!}{%
\begin{tabular}{|c|c|c|c|c|c|}
\hline
\multicolumn{2}{|c|}{\multirow{2}{*}{\textbf{QoS Attributes}}} & \multirow{2}{*}{\textbf{\begin{tabular}[c]{@{}c@{}}QoS\\ Unit\end{tabular}}} & \multirow{2}{*}{\textbf{\begin{tabular}[c]{@{}c@{}}Consumer\\ QoS Request\end{tabular}}} & \multicolumn{2}{c|}{\multirow{2}{*}{\textbf{\begin{tabular}[c]{@{}c@{}}Consumer\\ QoS Weights\end{tabular}}}} \\
\multicolumn{2}{|c|}{} &  &  & \multicolumn{2}{c|}{} \\ \hline
\textbf{Accountability} & \textbf{Levels} & \textbf{(1-10)} & 4 & 1.0 & = 1 \\ \hline
\multirow{6}{*}{\textbf{\begin{tabular}[c]{@{}c@{}}Agility\\ (Capacity)\end{tabular}}} & \multirow{3}{*}{\textbf{\begin{tabular}[c]{@{}c@{}}Number\\ of\\ vCPUs\end{tabular}}} & \multirow{3}{*}{\textbf{\begin{tabular}[c]{@{}c@{}}(4-core\\ each)\end{tabular}}} & \multirow{3}{*}{4} & \multirow{3}{*}{0.4} & \multirow{6}{*}{= 1} \\
 &  &  &  &  &  \\
 &  &  &  &  &  \\ \cline{2-5}
 & \textbf{vCPU Speed} & \textbf{(GHZ)} & 3.6 & 0.2 &  \\ \cline{2-5}
 & \textbf{Disk} & \textbf{(TB)} & 0.5 & 0.3 &  \\ \cline{2-5}
 & \textbf{Memory} & \textbf{(GB)} & 16 & 0.1 &  \\ \hline
\multirow{6}{*}{\textbf{Cost}} & \textbf{vCPU} & \textbf{(\$/h)} & 0.54 & 0.6 & \multirow{6}{*}{= 1} \\ \cline{2-5}
 & \multirow{4}{*}{\textbf{\begin{tabular}[c]{@{}c@{}}Data\\ Transfer\\ Bandwidth\end{tabular}}} & \multirow{2}{*}{\textbf{\begin{tabular}[c]{@{}c@{}}In\\ (\$/TB-m)\end{tabular}}} & \multirow{2}{*}{10} & \multirow{2}{*}{0.1} &  \\
 &  &  &  &  &  \\ \cline{3-5}
 &  & \multirow{2}{*}{\textbf{\begin{tabular}[c]{@{}c@{}}Out\\ (\$/TB-m)\end{tabular}}} & \multirow{2}{*}{51} & \multirow{2}{*}{0.1} &  \\
 &  &  &  &  &  \\ \cline{2-5}
 & \textbf{Storage} & \textbf{(\$/TB-m)} & 50 & 0.2 &  \\ \hline
\multirow{3}{*}{\textbf{Assurance}} & \multirow{2}{*}{\textbf{\begin{tabular}[c]{@{}c@{}}Support\\ (Levels)\end{tabular}}} & \multirow{2}{*}{\textbf{(1-10)}} & \multirow{2}{*}{8} & \multirow{2}{*}{0.3} & \multirow{3}{*}{= 1} \\
 &  &  &  &  &  \\ \cline{2-5}
 & \textbf{Availability} & \textbf{(\%)} & 99.9 & 0.7 &  \\ \hline
\textbf{Security} & \textbf{Levels} & \textbf{(1-10)} & 10 & 1.0 & = 1 \\ \hline
\multirow{2}{*}{\textbf{Satisfaction}} & \multirow{2}{*}{\textbf{\begin{tabular}[c]{@{}c@{}}Feedback\\ (Levels)\end{tabular}}} & \multirow{2}{*}{\textbf{(1-10)}} & \multirow{2}{*}{9} & \multirow{2}{*}{1.0} & \multirow{2}{*}{= 1} \\
 &  &  &  &  &  \\ \hline
\multirow{4}{*}{\textbf{Performance}} & \multirow{2}{*}{\textbf{\begin{tabular}[c]{@{}c@{}}Response\\ Time\end{tabular}}} & \multirow{2}{*}{\textbf{(sec)}} & \multicolumn{3}{c|}{\multirow{8}{*}{\textbf{}}} \\
 &  &  & \multicolumn{3}{c|}{} \\ \cline{2-3}
 & \multirow{2}{*}{\textbf{vCPU Speed}} & \multirow{2}{*}{\textbf{(GHZ)}} & \multicolumn{3}{c|}{} \\
 &  &  & \multicolumn{3}{c|}{} \\ \cline{1-3}
\multirow{4}{*}{\textbf{\begin{tabular}[c]{@{}c@{}}Network\\ Layer\\ QoS\end{tabular}}} & \multirow{2}{*}{\textbf{\begin{tabular}[c]{@{}c@{}}Down\\ Time\end{tabular}}} & \multirow{2}{*}{\textbf{(min)}} & \multicolumn{3}{c|}{} \\
 &  &  & \multicolumn{3}{c|}{} \\ \cline{2-3}
 & \textbf{Latency} & \textbf{(ms)} & \multicolumn{3}{c|}{} \\ \cline{2-3}
 & \textbf{Throughput} & \textbf{(mb/s)} & \multicolumn{3}{c|}{} \\ \hline
\end{tabular}%
}
\end{table}

\begin{figure}[ht]
    \centering
       {\includegraphics[width=.38\textheight, height=.24\textheight]{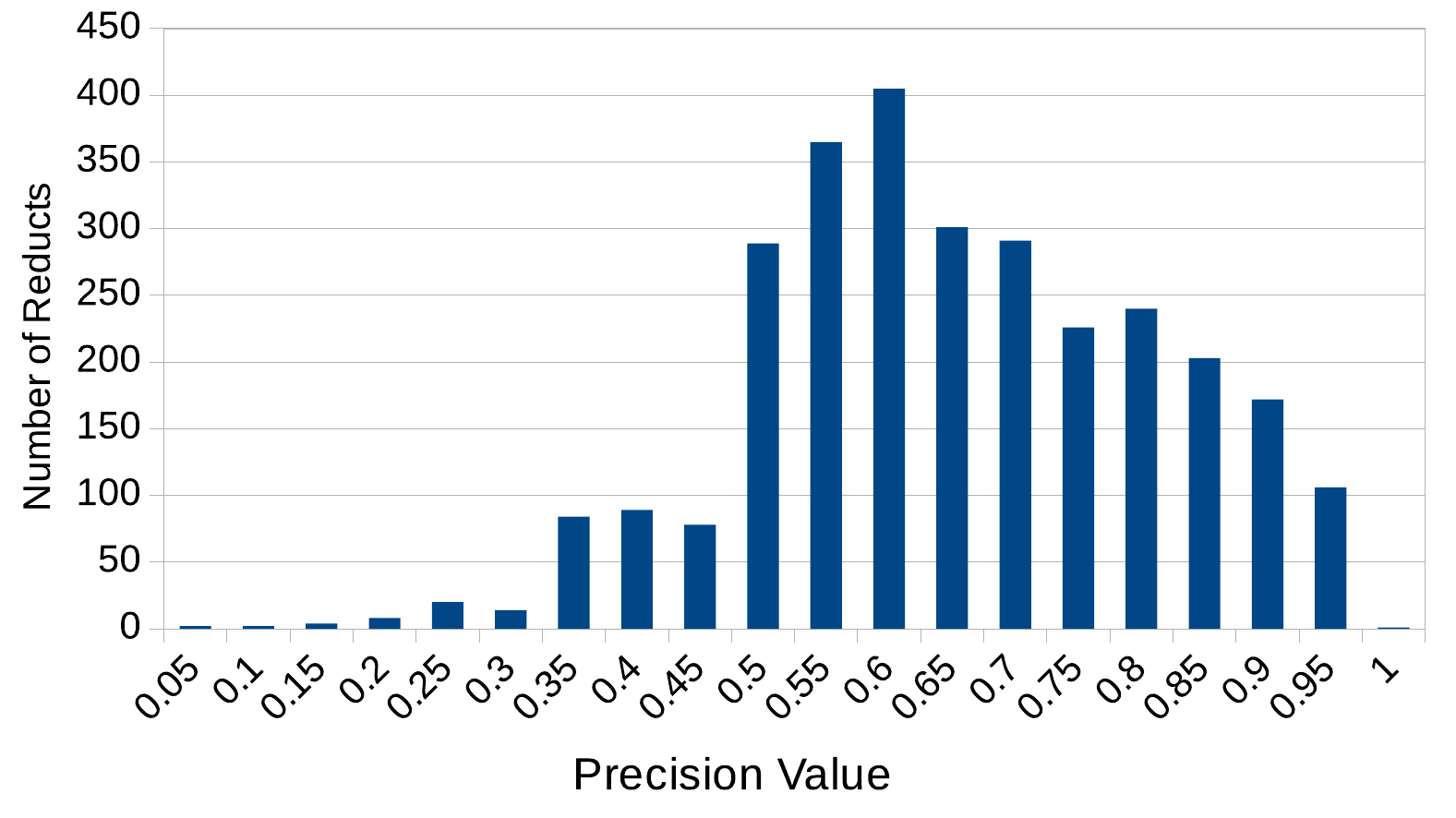}}\vspace{-.2cm}
    \caption{Precision Value vs Number of Reducts}
   \label{fig:precision}
\end{figure}

\begin{figure}[ht]
    \hspace{-.3cm}
       {\includegraphics[width=.37\textheight, height=.21\textheight]{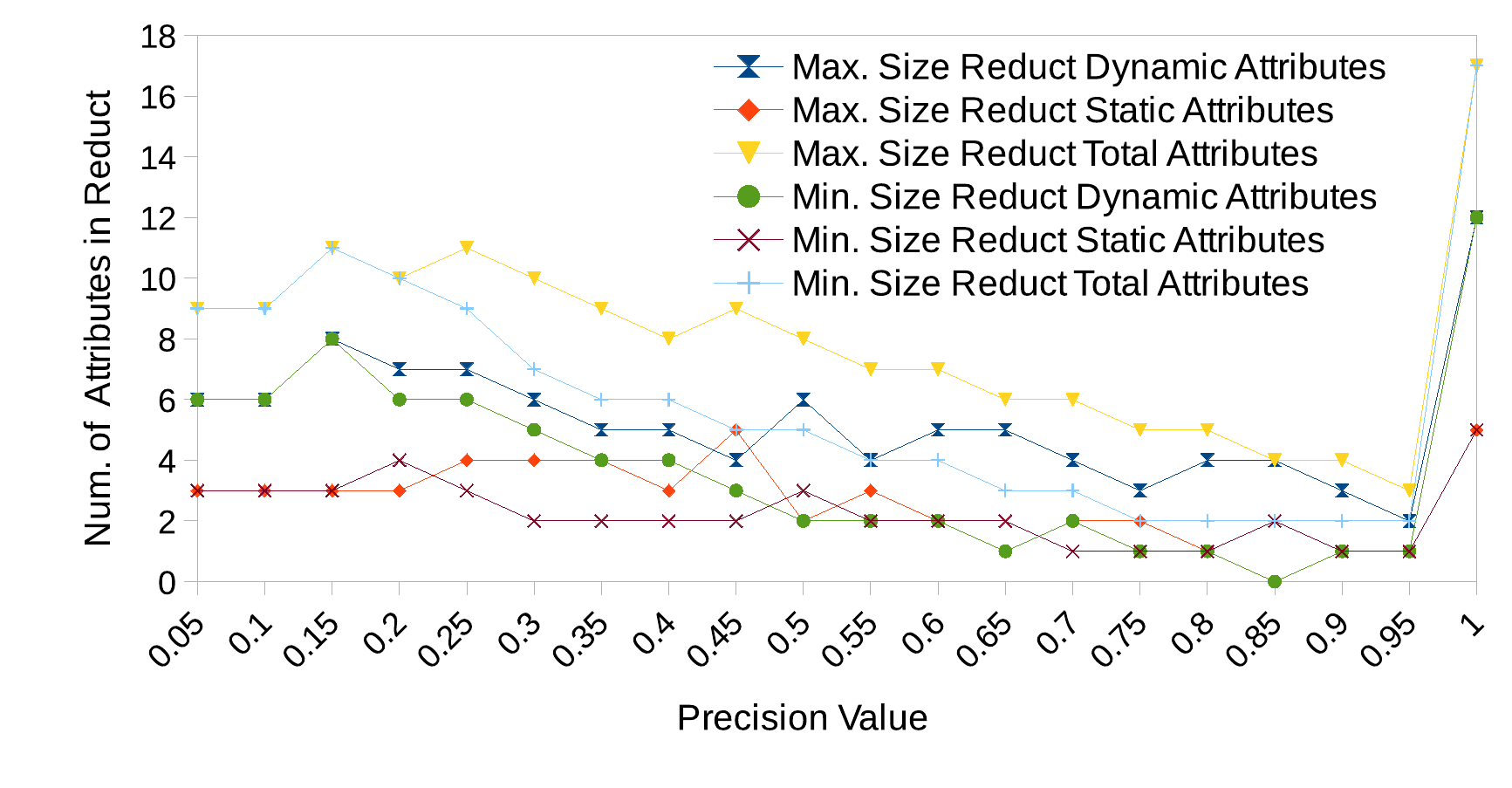}}\vspace{-.2cm}
    \caption{Precision Value vs Static and Dynamic Attributes in Maximum and Minimum Size QoS Reduct}
   \label{fig:precision1}
\end{figure}

In the third step, once DS is generated, search space reduction is achieved by employing all reduct concept of Fuzzy Rough Set Theory (R Language package RoughSet \cite{RoughSets}) $all\_reduct\_computation()$ function. This function gives all possible reducts of the DS, out of which one reduct is selected. Further, the number of reducts depend on the precision value ($\alpha$ precision), we also analyzed the precision value impact on the number of reducts as shown in Table \ref{tbl precisison} and in Figure \ref{fig:precision} \ref{fig:precision1}. With the change in precision value number of QoS attributes in reduct is decreases but the total number of reducts increases. For the experiment, we fixed the $\alpha$ precision $=$ $0.15$ as default value and got four reducts as shown in TABLE \ref{tbl:reduct}. Among four reducts the best reduct is selected using \textit{Best Reduct Algorithm} (Algorithm \ref{algo:BestReduct}). Here all four reducts consists of seven dynamic attributes, among these reducts anyone reduct can be selected. Based on selected reduct-1 Reduced Decision System (RDS) is generated and shown in TABLE \ref{RDS}. During the best reduct selection, Network Layer QoS attributes may not present in the reduct since user request does not consist this QoS attribute. The apparent reason for that can be that user does not have control over Network Layer QoS, it depends on the network traffic. Hence we added the Network Layer QoS in RDS if attributes are not part of RDS. A critical observation here is that in primary IS (TABLE \ref{infsys}) we have $17$ different quality attributes while after performing search space reduction based on selected best reduct and Network QoS attributes  (if not in RDS add them) we have only $11$ QoS attributes. In search space reduction the reduction of information we achieved is $35.29$\%. At present, more than 500 CSPs offering more than thousand different services (this static are based on the number of CSPs registered with Cloud Service Market \cite{Cloudservicemarket} and Cloud Harmony \cite{cloudharmony}). So if IS is vast (thousands of CSPs and a large number of QoS attributes), search space reduction will help a lot in better ranking of the CSPs. 

\begin{table}[ht]
\centering
\caption{All Possible Reducts of The Decision System}
\label{tbl:reduct}
\resizebox{.5\textwidth}{!}{%
\begin{tabular}{|c|c|c|c|c|c|}
\hline
\textbf{Reduct} & \textbf{1} & \textbf{2} & \textbf{3} & \textbf{4} &  \\ \hline
 & {\color[HTML]{036400} Accountability Level} & {\color[HTML]{036400} vCPU} & {\color[HTML]{036400} Accountability Level} & {\color[HTML]{036400} vCPU} & {\color[HTML]{036400} } \\ \cline{2-5}
 & {\color[HTML]{036400} vCPU Speed} & {\color[HTML]{036400} vCPU Speed} & {\color[HTML]{036400} vCPU Speed} & {\color[HTML]{036400} vCPU Speed} & {\color[HTML]{036400} } \\ \cline{2-5}
 & {\color[HTML]{036400} Memory} & {\color[HTML]{036400} Memory} & {\color[HTML]{036400} Memory} & {\color[HTML]{036400} Memory} & \multirow{-3}{*}{{\color[HTML]{036400} \begin{tabular}[c]{@{}c@{}}Green Color: \\ Dynamic\\ Attributes\end{tabular}}} \\ \cline{2-6} 
 & {\color[HTML]{9A0000} vCPU Cost} & {\color[HTML]{9A0000} vCPU Cost} & {\color[HTML]{9A0000} vCPU Cost} & {\color[HTML]{9A0000} vCPU Cost} & {\color[HTML]{9A0000} } \\ \cline{2-5}
 & {\color[HTML]{9A0000} In Bound Cost} & {\color[HTML]{9A0000} In Bound Cost} & {\color[HTML]{9A0000} Security Level} & {\color[HTML]{9A0000} Security Level} & {\color[HTML]{9A0000} } \\ \cline{2-5}
 & {\color[HTML]{9A0000} Out Bound Cost} & {\color[HTML]{9A0000} Out Bound Cost} & {\color[HTML]{9A0000} Out Bound Cost} & {\color[HTML]{9A0000} Out Bound Cost} & \multirow{-3}{*}{{\color[HTML]{9A0000} \begin{tabular}[c]{@{}c@{}}Red Color:\\ Static\\ Attribute\end{tabular}}} \\ \cline{2-6} 
 & {\color[HTML]{036400} Availability} & {\color[HTML]{036400} Availability} & {\color[HTML]{036400} Availability} & {\color[HTML]{036400} Availability} & {\color[HTML]{010066} } \\ \cline{2-5}
 & {\color[HTML]{036400} Response Time} & {\color[HTML]{036400} Response Time} & {\color[HTML]{036400} Response Time} & {\color[HTML]{036400} Response Time} & {\color[HTML]{010066} } \\ \cline{2-5}
 & {\color[HTML]{010066} Latency} & {\color[HTML]{010066} Latency} & {\color[HTML]{010066} Latency} & {\color[HTML]{010066} Latency} & {\color[HTML]{010066} } \\ \cline{2-5}
\multirow{-10}{*}{\textbf{\begin{tabular}[c]{@{}c@{}}QoS\\ Attributes\end{tabular}}} & {\color[HTML]{010066} Throughput} & {\color[HTML]{010066} Throughput} & {\color[HTML]{010066} Throughput} & {\color[HTML]{010066} Throughput} & \multirow{-4}{*}{{\color[HTML]{010066} \begin{tabular}[c]{@{}c@{}}Blue Color: \\ Network \\ QoS Attributes\\ (Dynamic)\end{tabular}}} \\ \hline
\end{tabular}%
}
\end{table}

\begin{table*}[ht]
\centering
\caption{Reduced Decision System (RDS) with QoS Weights}
\label{RDS}
\resizebox{\textwidth}{!}{%
\begin{tabular}{|c|c|c|c|c|c|c|c|c|c|c|c|c|l|c|c|}
\hline
\multicolumn{2}{|c|}{} &  & \multicolumn{10}{c|}{\textbf{Cloud Service Providers (CSPs)}} & \multicolumn{2}{c|}{} &  \\ \cline{4-13}
\multicolumn{2}{|c|}{\multirow{-2}{*}{\textbf{QoS Attributes}}} & \multirow{-2}{*}{\textbf{\begin{tabular}[c]{@{}c@{}}QoS\\ Unit\end{tabular}}} & \textbf{\begin{tabular}[c]{@{}c@{}}Google\\ Compute\\ Engine\end{tabular}} & \textbf{\begin{tabular}[c]{@{}c@{}}Storm\\ on\\ Demand\end{tabular}} & \textbf{\begin{tabular}[c]{@{}c@{}}Century\\ Link\end{tabular}} & \textbf{\begin{tabular}[c]{@{}c@{}}Amazon\\ EC2\end{tabular}} & \textbf{\begin{tabular}[c]{@{}c@{}}Vultr\\ Cloud\end{tabular}} & \textbf{\begin{tabular}[c]{@{}c@{}}IBM\\ Soft\\ Layer\end{tabular}} & \textbf{Linode} & \textbf{\begin{tabular}[c]{@{}c@{}}Digital\\ Ocean\end{tabular}} & \textbf{\begin{tabular}[c]{@{}c@{}}Micro-\\ soft\\ Azure\end{tabular}} & \textbf{\begin{tabular}[c]{@{}c@{}}Racks-\\ pace\end{tabular}} & \multicolumn{2}{c|}{\multirow{-2}{*}{\textbf{\begin{tabular}[c]{@{}c@{}}System QoS\\ Weights\\ (Level-1)\end{tabular}}}} & \multirow{-2}{*}{\textbf{\begin{tabular}[c]{@{}c@{}}Consumer\\ QoS Weights\\ (Level-2)\end{tabular}}} \\ \hline
\textbf{Accountability} & \textbf{Levels} & \textbf{(1-10)} & 8 & 4 & 4 & 9 & 3 & 7 & 1 & 6 & 8 & 2 & \textbf{0.095} &  & \textbf{1.0} \\ \cline{1-14} \cline{16-16} 
 & \textbf{vCPU Speed} & \textbf{(GHZ)} & 2.6 & 2.7 & 3.6 & 3.6 & 3.8 & 3.4 & 2.3 & 2.2 & 3.5 & 3.3 & \textbf{0.095} &  & \textbf{0.2} \\ \cline{2-14} \cline{16-16} 
\multirow{-2}{*}{\textbf{\begin{tabular}[c]{@{}c@{}}Agility\\ (Capacity)\end{tabular}}} & \textbf{Memory} & \textbf{(GB)} & 14.4 & 16 & 16 & 15 & 24 & 32 & 16 & 16 & 32 & 15 & \textbf{0.095} &  & \textbf{0.1} \\ \cline{1-14} \cline{16-16} 
 & \textbf{vCPU} & \textbf{(\$/h)} & 0.56 & 0.48 & 0.56 & 0.39 & 0.44 & 0.69 & 0.48 & 0.23 & 0.38 & 0.51 & \textbf{0.095} &  & \textbf{0.6} \\ \cline{2-14} \cline{16-16} 
 &  &  &  &  &  &  &  &  &  &  &  &  &  &  &  \\
 &  & \multirow{-2}{*}{\textbf{\begin{tabular}[c]{@{}c@{}}In\\ (\$/TB-m)\end{tabular}}} & \multirow{-2}{*}{0} & \multirow{-2}{*}{8} & \multirow{-2}{*}{10} & \multirow{-2}{*}{0} & \multirow{-2}{*}{7} & \multirow{-2}{*}{8} & \multirow{-2}{*}{10} & \multirow{-2}{*}{9} & \multirow{-2}{*}{18.43} & \multirow{-2}{*}{15.2} & \multirow{-2}{*}{\textbf{0.095}} &  & \multirow{-2}{*}{\textbf{0.1}} \\ \cline{3-14} \cline{16-16} 
 &  &  &  &  &  &  &  &  &  &  &  &  &  &  &  \\
\multirow{-5}{*}{\textbf{Cost}} & \multirow{-4}{*}{\textbf{\begin{tabular}[c]{@{}c@{}}Data\\ Transfer\\ Bandwidth\end{tabular}}} & \multirow{-2}{*}{\textbf{\begin{tabular}[c]{@{}c@{}}Out\\ (\$/TB-m)\end{tabular}}} & \multirow{-2}{*}{70.6} & \multirow{-2}{*}{40} & \multirow{-2}{*}{51.2} & \multirow{-2}{*}{51.2} & \multirow{-2}{*}{22.86} & \multirow{-2}{*}{92.16} & \multirow{-2}{*}{51.2} & \multirow{-2}{*}{45.72} & \multirow{-2}{*}{18.43} & \multirow{-2}{*}{40} & \multirow{-2}{*}{\textbf{0.095}} &  & \multirow{-2}{*}{\textbf{0.1}} \\ \cline{1-14} \cline{16-16} 
\textbf{Assurance} & \textbf{Availability} & \textbf{(\%)} & 99.99 & 100 & 99.97 & 99.99 & 99.89 & 99.97 & 100 & 99.99 & 100 & 99.95 & \textbf{0.095} & \multirow{-9}{*}{\textbf{\begin{tabular}[c]{@{}c@{}}0.095\\ *\\ 7\\ =\\ 0.67\end{tabular}}} & \textbf{0.1} \\ \hline
\textbf{Performance} & \textbf{Response Time} & \textbf{(sec)} & 83.5 & 90 & 97 & 52 & 90 & 100 & 97 & 85 & 76 & 57 & \textbf{0.082} & \multicolumn{2}{c|}{} \\ \cline{1-14}
 &  &  &  &  &  &  &  &  &  &  &  &  &  & \multicolumn{2}{c|}{} \\
 & \multirow{-2}{*}{\textbf{\begin{tabular}[c]{@{}c@{}}Down\\ Time\end{tabular}}} & \multirow{-2}{*}{\textbf{(min)}} & \multirow{-2}{*}{1.02} & \multirow{-2}{*}{0} & \multirow{-2}{*}{1.98} & \multirow{-2}{*}{0.51} & \multirow{-2}{*}{9.05} & \multirow{-2}{*}{7.5} & \multirow{-2}{*}{0} & \multirow{-2}{*}{1.53} & \multirow{-2}{*}{2.83} & \multirow{-2}{*}{2.83} & \multirow{-2}{*}{\textbf{0.082}} & \multicolumn{2}{c|}{} \\ \cline{2-14}
 & \textbf{Latency} & \textbf{(ms)} & 31 & 57 & 31 & 29 & 28 & 29 & 28 & 28 & 32 & 32 & \textbf{0.082} & \multicolumn{2}{c|}{} \\ \cline{2-14}
\multirow{-4}{*}{\textbf{\begin{tabular}[c]{@{}c@{}}Network\\ Layer\\ QoS\end{tabular}}} & \textbf{Throughput} & \textbf{(mb/s)} & 20.24 & 16.99 & 24.99 & 16.23 & 10.11 & 16.23 & 8.12 & 24.67 & 23.11 & 23.67 & \textbf{0.082} & \multicolumn{2}{c|}{\multirow{-5}{*}{\textbf{\begin{tabular}[c]{@{}c@{}}0.082 * 4\\ =\\ 0.33\end{tabular}}}} \\ \hline
\multicolumn{3}{|c|}{{\color[HTML]{CB0000} \textbf{Decision Attribute}}} & {\color[HTML]{CB0000} \textbf{3}} & {\color[HTML]{CB0000} \textbf{1}} & {\color[HTML]{CB0000} \textbf{3}} & {\color[HTML]{CB0000} \textbf{2}} & {\color[HTML]{CB0000} \textbf{1}} & {\color[HTML]{CB0000} \textbf{3}} & {\color[HTML]{CB0000} \textbf{1}} & {\color[HTML]{CB0000} \textbf{2}} & {\color[HTML]{CB0000} \textbf{2}} & {\color[HTML]{CB0000} \textbf{2}} & \multicolumn{3}{c|}{{\color[HTML]{CB0000} \textbf{-}}} \\ \hline
\end{tabular}%
}
\end{table*}

In the fourth step, once RDS is available, two-level weight assignment to QoS attribute is performed (as explained earlier in Section \ref{BRM}). After weight assignment, by using User QoS Request (as shown in TABLE \ref{tbl:cqos}) and RDS (TABLE \ref{RDS}), Weighted Euclidean Distance (Score) of respective CSP is computed. Based on the Score CSPs are sorted in ascending order where smaller score represents better CSPs by concerning user QoS request (TABLE \ref{ranklist}). Hence, the corresponding ranking of all the CSPs can be determined based on Score (Weighted Euclidean Distance) (4.989, 5.404, 6.843, 7.372, 7.387, 7.916, 8.403, 8.450, 8.668, 8.814). The ranked list of the CSPs is as follows: \textit{Amazon EC2 $>$ Rackspace $>$ Microsoft Azure $>$ Google Compute Engine $>$ Digital Ocean $>$ Vultr Cloud $>$ Century Link $>$ IBM Soft Layer $>$ Storm on Demand}. Finally, based on user QoS request \textit{Amazon EC2} give the best service. In next step, a ranked list of CSPs is sent to the user for service selection. Finally, the user selects a CSP and gives his selection response to the system, system communicated with selected service provider for resource reservation and service execution. During service execution we monitor the execution based on SLA to improve the accuracy and use the monitoring data in designing IS for next ranking process (as shown in Figure \ref{fig:RC}).

\begin{table*}[ht]
\centering
\caption{Normalized Weighted Reduced Decision System and Weighted Euclidean Distance}
\label{ranklist}
\resizebox{\textwidth}{!}{%
\begin{tabular}{|c|c|c|c|c|c|c|c|c|c|c|c|c|}
\hline
\multicolumn{2}{|c|}{\multirow{4}{*}{\textbf{\begin{tabular}[c]{@{}c@{}}QoS\\ Attributes\end{tabular}}}} & \multirow{4}{*}{\textbf{Unit}} & \multicolumn{10}{c|}{\textbf{Cloud Service Providers (CSPs)}} \\ \cline{4-13} 
\multicolumn{2}{|c|}{} &  & \multirow{3}{*}{\textbf{\begin{tabular}[c]{@{}c@{}}Amazon\\ EC2\end{tabular}}} & \multirow{3}{*}{\textbf{Rackspace}} & \multirow{3}{*}{\textbf{\begin{tabular}[c]{@{}c@{}}Microsoft\\ Azure\end{tabular}}} & \multirow{3}{*}{\textbf{\begin{tabular}[c]{@{}c@{}}Google\\ Compute\\ Engine\end{tabular}}} & \multirow{3}{*}{\textbf{\begin{tabular}[c]{@{}c@{}}Digital\\ Ocean\end{tabular}}} & \multirow{3}{*}{\textbf{\begin{tabular}[c]{@{}c@{}}Vultr\\ Cloud\end{tabular}}} & \multirow{3}{*}{\textbf{\begin{tabular}[c]{@{}c@{}}Century\\ Link\end{tabular}}} & \multirow{3}{*}{\textbf{Linode}} & \multirow{3}{*}{\textbf{\begin{tabular}[c]{@{}c@{}}IBM\\ Soft\\ Layer\end{tabular}}} & \multirow{3}{*}{\textbf{\begin{tabular}[c]{@{}c@{}}Storm\\ on\\ Demand\end{tabular}}} \\
\multicolumn{2}{|c|}{} &  &  &  &  &  &  &  &  &  &  &  \\
\multicolumn{2}{|c|}{} &  &  &  &  &  &  &  &  &  &  &  \\ \hline
\textbf{Accountability} & \textbf{Levels} & \textbf{(1-10)} & 0.861 & 0.191 & 0.766 & 0.766 & 0.574 & 0.287 & 0.383 & 0.096 & 0.670 & 0.383 \\ \hline
\multirow{2}{*}{\textbf{\begin{tabular}[c]{@{}c@{}}Agility\\ (Capacity)\end{tabular}}} & \textbf{vCPU Speed} & \textbf{(GHZ)} & 0.069 & 0.063 & 0.067 & 0.050 & 0.042 & 0.073 & 0.069 & 0.044 & 0.065 & 0.052 \\ \cline{2-13} 
 & \textbf{Memory} & \textbf{(GB)} & 0.431 & 0.431 & 0.919 & 0.413 & 0.459 & 0.689 & 0.459 & 0.459 & 0.919 & 0.459 \\ \hline
\multirow{5}{*}{\textbf{Cost}} & \textbf{vCPU} & \textbf{(\$/h)} & 0.022 & 0.029 & 0.022 & 0.032 & 0.013 & 0.025 & 0.032 & 0.028 & 0.040 & 0.028 \\ \cline{2-13} 
 & \multirow{4}{*}{\textbf{\begin{tabular}[c]{@{}c@{}}Data\\ Transfer\\ Bandwidth\end{tabular}}} & \multirow{2}{*}{\textbf{\begin{tabular}[c]{@{}c@{}}In\\ (\$/TB-m)\end{tabular}}} & \multirow{2}{*}{0.000} & \multirow{2}{*}{0.145} & \multirow{2}{*}{0.176} & \multirow{2}{*}{0.000} & \multirow{2}{*}{0.086} & \multirow{2}{*}{0.067} & \multirow{2}{*}{0.096} & \multirow{2}{*}{0.096} & \multirow{2}{*}{0.077} & \multirow{2}{*}{0.077} \\
 &  &  &  &  &  &  &  &  &  &  &  &  \\ \cline{3-13} 
 &  & \multirow{2}{*}{\textbf{\begin{tabular}[c]{@{}c@{}}Out\\ (\$/TB-m)\end{tabular}}} & \multirow{2}{*}{0.490} & \multirow{2}{*}{0.383} & \multirow{2}{*}{0.176} & \multirow{2}{*}{0.676} & \multirow{2}{*}{0.438} & \multirow{2}{*}{0.219} & \multirow{2}{*}{0.490} & \multirow{2}{*}{0.490} & \multirow{2}{*}{0.882} & \multirow{2}{*}{0.383} \\
 &  &  &  &  &  &  &  &  &  &  &  &  \\ \hline
\textbf{Assurance} & \textbf{Availability} & \textbf{(\%)} & 6.699 & 6.697 & 6.700 & 6.699 & 6.699 & 6.693 & 6.698 & 6.700 & 6.698 & 6.700 \\ \hline
\multirow{2}{*}{\textbf{Performance}} & \multirow{2}{*}{\textbf{\begin{tabular}[c]{@{}c@{}}Response\\ Time\end{tabular}}} & \multirow{2}{*}{\textbf{(sec)}} & \multirow{2}{*}{4.290} & \multirow{2}{*}{4.703} & \multirow{2}{*}{6.270} & \multirow{2}{*}{6.889} & \multirow{2}{*}{7.013} & \multirow{2}{*}{7.425} & \multirow{2}{*}{8.003} & \multirow{2}{*}{8.003} & \multirow{2}{*}{8.250} & \multirow{2}{*}{7.425} \\
 &  &  &  &  &  &  &  &  &  &  &  &  \\ \hline
\multirow{3}{*}{\textbf{\begin{tabular}[c]{@{}c@{}}Network\\ Layer\\ QoS\end{tabular}}} & \textbf{Down Time} & \textbf{(min)} & 0.042 & 0.233 & 0.233 & 0.084 & 0.126 & 0.747 & 0.163 & 0.000 & 0.619 & 0.000 \\ \cline{2-13} 
 & \textbf{Latency} & \textbf{(ms)} & 2.393 & 2.640 & 2.640 & 2.558 & 2.310 & 2.310 & 2.558 & 2.310 & 2.393 & 4.703 \\ \cline{2-13} 
 & \textbf{Throughput} & \textbf{(mb/s)} & 1.339 & 1.953 & 1.907 & 1.670 & 2.035 & 0.834 & 2.062 & 0.670 & 1.339 & 1.402 \\ \hline
\multicolumn{3}{|c|}{\multirow{2}{*}{\textbf{Weighted Euclidean Distance (Score)}}} & \multirow{2}{*}{\textbf{4.989}} & \multirow{2}{*}{\textbf{5.404}} & \multirow{2}{*}{\textbf{6.843}} & \multirow{2}{*}{\textbf{7.372}} & \multirow{2}{*}{\textbf{7.387}} & \multirow{2}{*}{\textbf{7.916}} & \multirow{2}{*}{\textbf{8.403}} & \multirow{2}{*}{\textbf{8.450}} & \multirow{2}{*}{\textbf{8.668}} & \multirow{2}{*}{\textbf{8.814}} \\
\multicolumn{3}{|c|}{} &  &  &  &  &  &  &  &  &  &  \\ \hline
\multicolumn{3}{|c|}{\textbf{Service Provider Rank}} & \textbf{1} & \textbf{2} & \textbf{3} & \textbf{4} & \textbf{5} & \textbf{6} & \textbf{7} & \textbf{8} & \textbf{9} & \textbf{10} \\ \hline
\end{tabular}%
}
\end{table*}

\begin{table*}[ht]
\centering
\caption{Change in the Number of Reducts, Static, and Dynamic Attributes in Maximum and Minimum QoS Reduct with Change in Precision Value ($0<=$ $\alpha$ $<=1$).}
\label{tbl precisison}
\resizebox{\textwidth}{!}{%
\begin{tabular}{|c|c|c|c|c|c|c|c|c|c|c|c|c|c|c|c|c|c|c|c|c|c|}
\hline
 
\multicolumn{2}{|c|}{\textbf{Precision Value}} & \textbf{0.05} & \textbf{0.1} & \textbf{0.15} & \textbf{0.2} & \textbf{0.25} & \textbf{0.3} & \textbf{0.35} & \textbf{0.4} & \textbf{0.45} & \textbf{0.5} & \textbf{0.55} & \textbf{0.6} & \textbf{0.65} & \textbf{0.7} & \textbf{0.75} & \textbf{0.8} & \textbf{0.85} & \textbf{0.9} & \textbf{0.95} & \textbf{1} \\ \hline
\multicolumn{2}{|c|}{\textbf{Number of Reducts}} & 2 & 2 & 4 & 8 & 20 & 14 & 84 & 89 & 78 & 289 & 365 & 405 & 301 & 291 & 226 & 240 & 203 & 172 & 106 & 1 \\ \hline
 & \textbf{Dynamic Attributes} & 6 & 6 & 8 & 7 & 7 & 6 & 5 & 5 & 4 & 6 & 4 & 5 & 5 & 4 & 3 & 4 & 4 & 3 & 2 & 12 \\ \cline{2-22} 
 & \textbf{Static Attributes} & 3 & 3 & 3 & 3 & 4 & 4 & 4 & 3 & 5 & 2 & 3 & 2 & 1 & 2 & 2 & 1 & 0 & 1 & 1 & 5 \\ \cline{2-22} 
\multirow{-3}{*}{\textbf{\begin{tabular}[c]{@{}c@{}}Reduct With\\ Maximum\\ No. of QoS\end{tabular}}} & \textbf{Total Attributes} & 9 & 9 & 11 & 10 & 11 & 10 & 9 & 8 & 9 & 8 & 7 & 7 & 6 & 6 & 5 & 5 & 4 & 4 & 3 & 17 \\ \hline
 &\textbf{Dynamic Attributes} & 6 & 6 & 8 & 6 & 6 & 5 & 4 & 4 & 3 & 2 & 2 & 2 & 1 & 2 & 1 & 1 & 0 & 1 & 1 & 12 \\ \cline{2-22} 
 & \textbf{Static Attributes} & 3 & 3 & 3 & 4 & 3 & 2 & 2 & 2 & 2 & 3 & 2 & 2 & 2 & 1 & 1 & 1 & 2 & 1 & 1 & 5 \\ \cline{2-22} 
\multirow{-3}{*}{\textbf{\begin{tabular}[c]{@{}c@{}}Reduct With\\ Minimum\\ No. of QoS\end{tabular}}} &\textbf{Total Attributes} & 9 & 9 & 11 & 10 & 9 & 7 & 6 & 6 & 5 & 5 & 4 & 4 & 3 & 3 & 2 & 2 & 2 & 2 & 2 & 17 \\ \hline
\end{tabular}%
}
\end{table*}

\begin{figure*}[]
    \centering
       {\includegraphics[width=.75\textheight, height=.43\textheight]{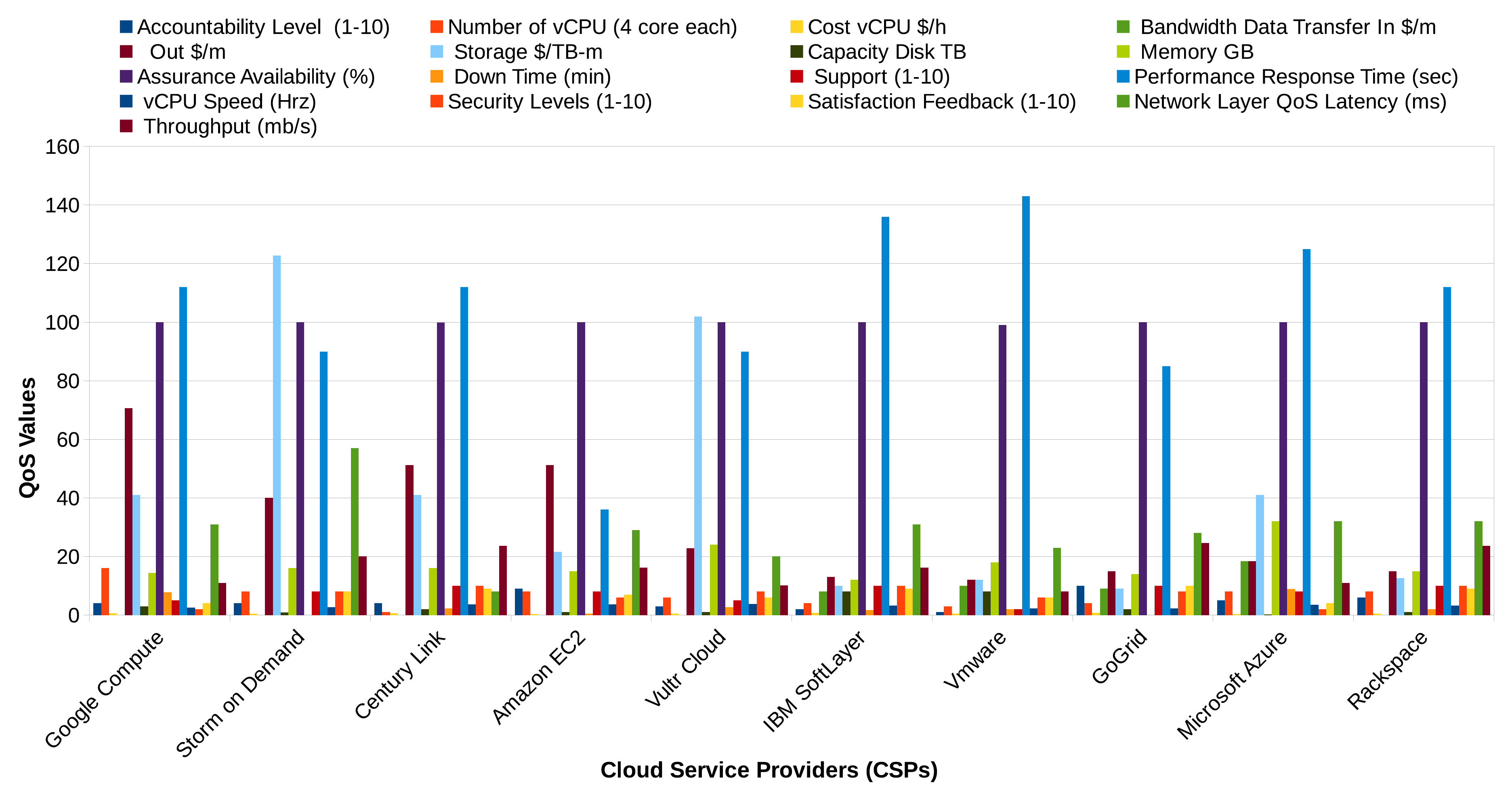}}\vspace{-.2cm}
    \caption{Cloud Service Provider QoS Values}
   \label{fig:result0}
\end{figure*}

\begin{figure}[ht]
    \centering
      {\includegraphics[width=.37\textheight, height=.27\textheight]{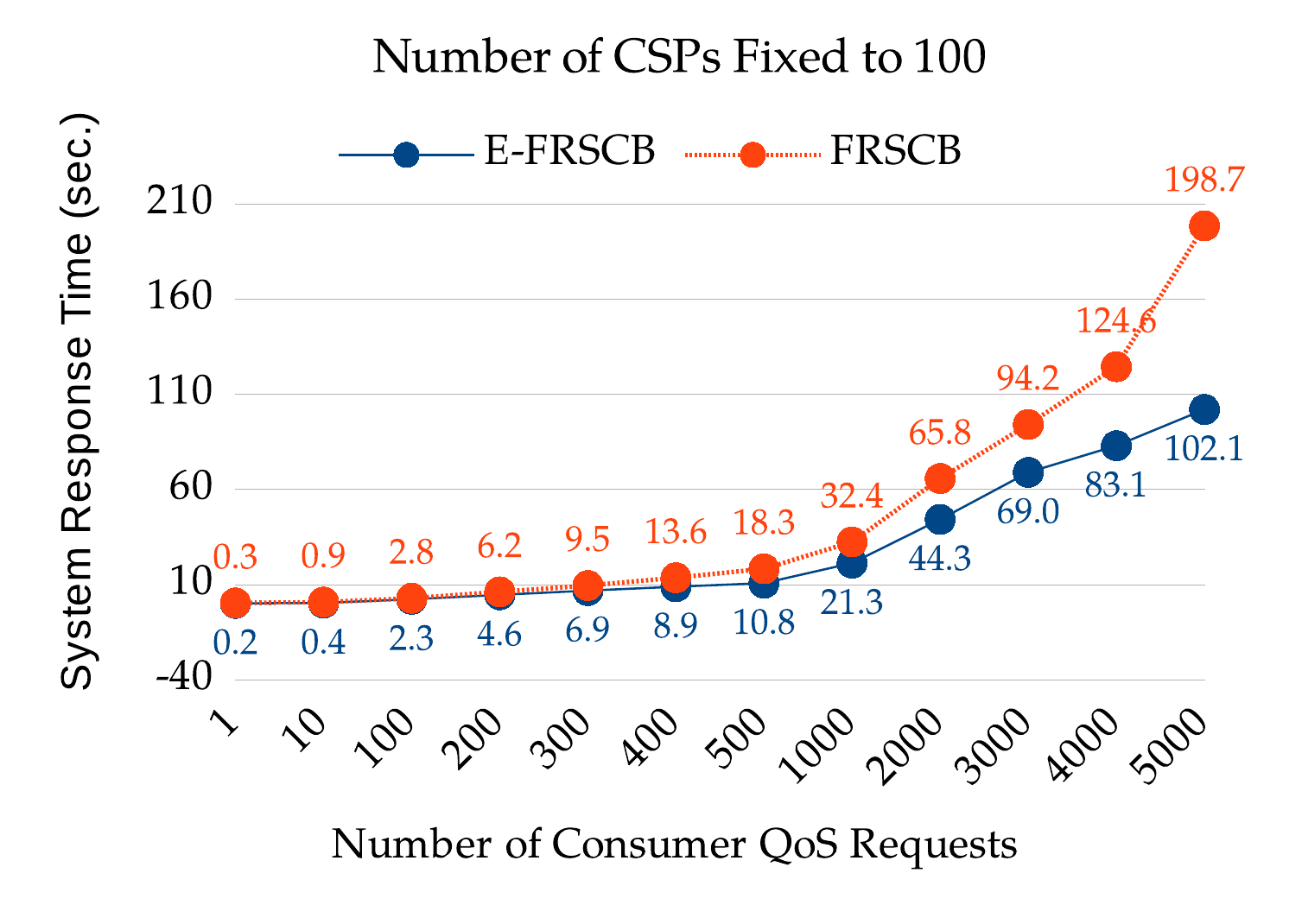}}\vspace{-.2cm}
    \caption{Number of QoS Request vs System Response Time}
   \label{fig:result}
\end{figure}

\begin{figure}[ht]
    \centering
       {\includegraphics[width=.37\textheight, height=.27\textheight]{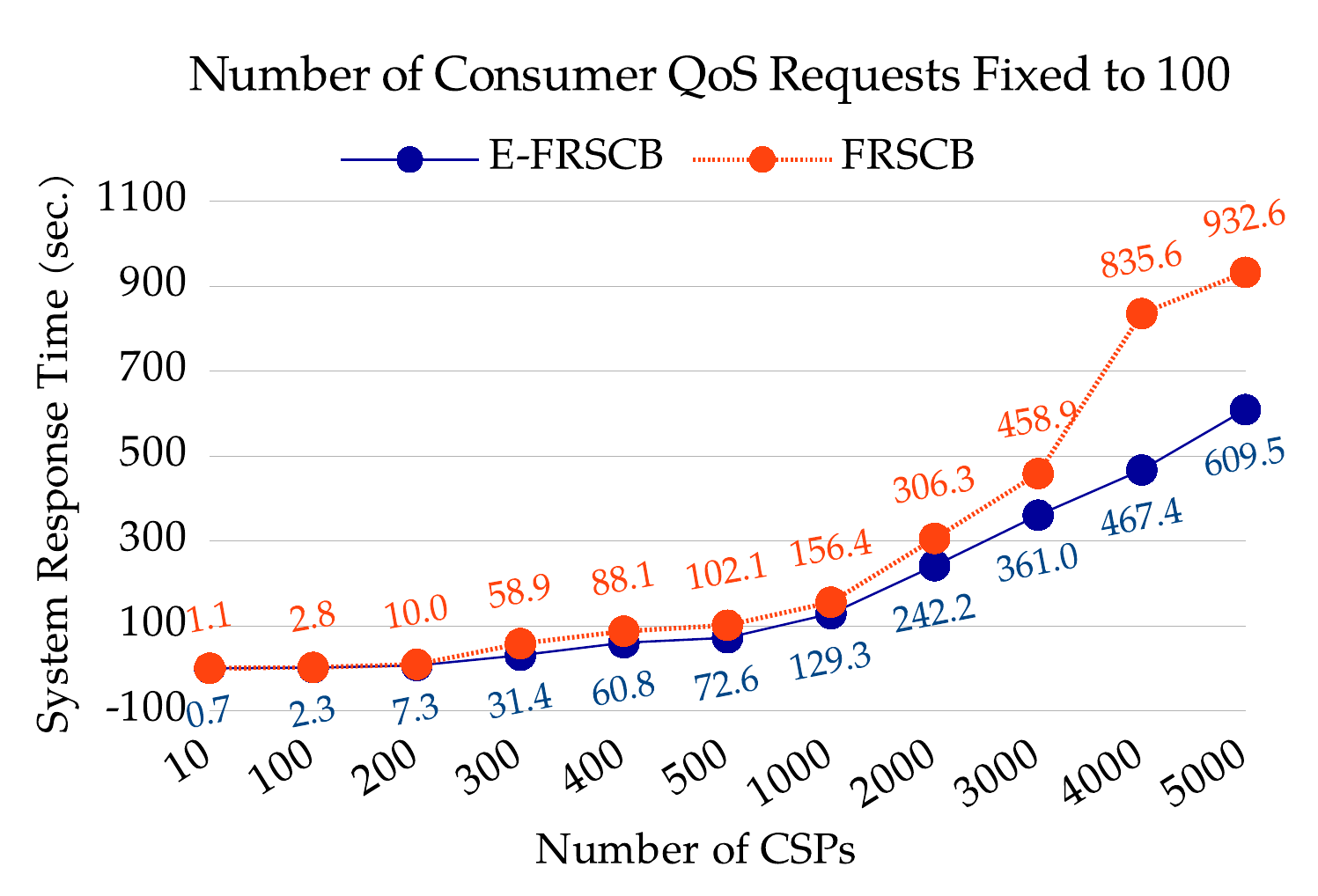}}\vspace{-.2cm}
    \caption{Number of CSPs vs System Response Time}
   \label{fig:result1}
\end{figure}

Two different experiments are performed to analyze the performance of the proposed technique with existing fuzzy rough set-based technique (FRSCB \cite{Anjana2017}). To simulate Compute cloud service infrastructure we used CloudSim \cite{cloudsim}. During simulation for the first test, we kept the number of CSPs to fix $100$ and submitted $1$ to $5000$ number of the user request for CSP ranking. In second experiment user, QoS request was set to $100$, and the number of CSPs ranges from $10$ to $5000$. For simulation, random QoS values are generated to design the IS based on the domain of each QoS attribute. For user request also QoS values and weights are assigned randomly. Furthermore using the information available in IS, service providers were created where each CSP consists of time shared Virtual Machine Scheduler and 50 computing hosts. During simulation for each task execution time is selected randomly from $0.1$ to $1.0$ ms. Finally based on both the experiment set up the response time (in the sec.) of proposed and FRSCB  ranking technique is recorded. Experiment result shows (Figure \ref{fig:result}, \ref{fig:result1}) that proposed method outperforms then FRSCB ranking technique. Here, Figure \ref{fig:result}, \ref{fig:result1} shows that our proposed architecture is scalable when both number user QoS request increases and also when the number of CSPs increases.

\section{Conclusions and Future Work}
\label{conclusion}
Because of the vast diversity of available cloud service from the user point of view, it leads to many challenges for discovering and managing their desired cloud services. As cloud service discovery and management involve various operational aspects, it is desired to have a cloud service broker who can do this task on behalf of the user. In this paper, we have presented an efficient Cloud Service Provider evaluation system using the Fuzzy Rough Set technique and presented a case study on IaaS service provider ranking. Our proposed architecture not only rank the cloud services but also monitor execution. The significant contributions of proposed brokerage system can be summarized as follows: 1) it evaluates the number of cloud service providers based on user QoS requirements and offers an opportunity to the user to select the best cloud service from a ranked list of the CSPs. 2) it also priorities the user request by incorporating user assign weights and gives the relative priority to non-user requested QoS by considering system assigned weights. 3) the primary focus was on search space reduction so that we can minimize the searching time and also improve the efficiency of the ranking procedure. 4) we used Weighted Euclidean Distance to lead to the ideal value (i.e., zero line) it shows the improved representation of the method. 5) finally we monitor the service execution once the selection is made by the users to get the historical data and actual performance of the cloud services to improve the accuracy of next service provider ranking process. The proposed approach can deal with hybrid information system and also scalable and efficient. This technique helps new users and the brokerage based organization to directly deal with fuzzy information system with there rough QoS requirement for CSPs ranking, and selection. In future, we are working to develop an online model by dynamically fetching all QoS attributes for service provider ranking and selection based on user QoS requirements.

\bibliographystyle{ieeetr}
\bibliography{Arxiv-EFRSCB-IEEET-V2}

\end{document}